\documentclass[12pt]{article}

\pdfoutput=1

\usepackage{jheppub}
\usepackage{graphicx,epsfig,amsmath,amssymb,booktabs}
\usepackage{caption,subcaption}
\usepackage{xspace}

\newlength{\unitcharwidth}
\newlength{\unitsuperscriptwidth}
\newcommand{\rd}{\ensuremath{\mathrm{d}}}

\newcommand{\Gmu}{\ensuremath{G_\mu}}
\newcommand{\alphazero}{\ensuremath{\alpha(0)}}

\newcommand{\alphas}{\ensuremath{\alpha_s}}
\newcommand{\aaa}{\ensuremath{\gamma\gamma\gamma}}
\newcommand{\aaw}{\ensuremath{\gamma\gamma W}}
\newcommand{\aaz}{\ensuremath{\gamma\gamma Z}}
\newcommand{\shortequal}{\ensuremath{\!\!\!=\!\!\!}}
\newcommand{\pT}{\ensuremath{p_\text{T}}}
\newcommand{\pTveto}{\ensuremath{p_\text{T,jet}^\text{veto}}}
\newcommand{\order}{\ensuremath{\mathcal{O}}}
\newcommand{\vP}{\ensuremath{\vphantom{\int_a^b}}}

\newcommand{\Sherpa}{S\scalebox{0.8}{HERPA}\xspace}
\newcommand{\GoSam}{G\scalebox{0.8}{O}S\scalebox{0.8}{AM}\xspace}

\newcommand{\QGraf}{QG\scalebox{0.8}{RAF}\xspace}
\newcommand{\FORM}{F\scalebox{0.8}{ORM}\xspace}
\newcommand{\Spinney}{S\scalebox{0.8}{PINNEY}\xspace}
\newcommand{\Ninja}{N\scalebox{0.8}{INJA}\xspace}
\newcommand{\Samurai}{S\scalebox{0.8}{AMURAI}\xspace}
\newcommand{\GolemNF}{G\scalebox{0.8}{OLEM}95\xspace}
\newcommand{\OneLoop}{O\scalebox{0.8}{NE}L\scalebox{0.8}{OOP}\xspace}

\newcommand{\QCDpEW}{QCD+EW}
\newcommand{\QCDtEW}{QCD$\times$EW}
\newcommand{\deltaQCD}{\ensuremath{\delta_\text{QCD}}}
\newcommand{\deltaEW}{\ensuremath{\delta_\text{EW}}}
\newcommand{\muR}{\ensuremath{\mu_{\mathrm{R}}}}
\newcommand{\muF}{\ensuremath{\mu_{\mathrm{F}}}}

\title{NLO QCD+EW corrections to diphoton production in association with a vector boson}

\author[a]{Nicolas Greiner,}
\author[a,b]{Marek Sch\"onherr}

\affiliation[a]{Physik Institut, Universit{\"a}t Z{\"u}rich, Winterthurerstr.190, 8057 Z\"urich, Switzerland}
\affiliation[b]{Theoretical Physics Department, CERN, 1211 Geneva 23, Switzerland}

\preprint{
  \small
  \begin{flushright}
    ZU--TH 31/17\\ CERN-TH-2017-226
  \end{flushright}
}

\abstract{
Processes with three external electroweak gauge boson allow for a measurement of triple and quartic gauge
couplings. They can be used to constrain anomalous gauge couplings, where new physics might predominantly
couple to electroweak gauge bosons. In this paper we chose a class of such processes where we consider two photons
and an additional vector boson in the final state. As additional vector boson we consider either a third photon or a
$W$ or $Z$ boson. For the latter two cases we assume a leptonic decay of the boson. We calculate the next-to-leading
order QCD and electroweak corrections to these processes with a particular emphasis on the until now unknown 
electroweak corrections. We find that the electroweak corrections to the total cross section are moderate in the range of a few 
per cent at most, but can reach several tens of per cent in regions of phase space  that are particularly interesting
 in the context of new physics searches. In addition we investigate the difference between additive and multiplicative scheme 
 when combining
QCD and electroweak corrections and we assess the importance of photon induced contributions to these processes. 
}


\hypersetup{
  pdftitle = {NLO QCD+EW corrections to diphoton production in association with a vector boson},
  pdfauthor = {Nicolas Greiner, Marek Schoenherr}
}

\begin{document}

\maketitle

\section{Introduction}
\label{sec:intro}

The precise understanding of the electroweak symmetry breaking 
mechanism is an important cornerstone of the LHC physics 
programme. 
Different realisations of the electroweak sector or new physics 
coupling to electroweak gauge bosons will yield to deviations 
compared to the Standard Model prediction. 
New physics effects can conveniently be described in terms of 
an effective theory where new heavy degrees of freedom are 
integrated out and deviations from the Standard Model are 
parametrised by higher dimensional operators 
\cite{Weinberg:1978kz,Weinberg:1979pi}. 
Higher dimensional operators can lead to deviations in triple 
and quartic gauge couplings. Moreover, new vertices 
(e.g. $\gamma\gamma\gamma$, $Z\gamma\gamma$) that do not 
exist in the Standard Model can appear. 
The class of processes that involve a pair of photons in 
association with another vector boson allows to measure 
deviations in triple and quartic gauge couplings and is 
therefore a particularly interesting class of processes. 
Consequently both ATLAS and CMS have measured these types of 
processes and derived constraints on anomalous gauge couplings 
\cite{Aad:2016sau,Sirunyan:2017lvq,Aad:2015uqa,Aad:2015bua}. 

In this paper we calculate the next-to-leading order QCD and 
electroweak corrections to the processes $\gamma\gamma\gamma$, 
$\gamma\gamma e^{+} e^{-}$ and $\gamma\gamma e^{-} \bar{\nu_e}$. 
For the latter two, all possible off-shell contributions of 
intermediate vector bosons are taken into account. 
The QCD corrections to these processes have already appeared 
in the literature 
\cite{Bozzi:2011en,Campbell:2012ft,Bozzi:2011wwa}. 
For a more complete picture of the higher order effects we 
recompute them for a centre of mass energy of 
13\,TeV, and supplement them with the next-to-leading order 
electroweak corrections. 
Although the electroweak corrections are much smaller at the level 
of the total cross section compared to the NLO QCD corrections 
they are particularly important when deriving limits on 
anomalous couplings. 
The effects of higher dimensional operators increase in the 
high energy tails of differential distributions and lead to a 
change of the shapes. And it is in the same region where the 
Sudakov logarithms from the electroweak corrections will play 
an essential role as well.

The paper is organized as follows. 
In Section \ref{sec:setup} we describe the calculational setup 
that has been used to obtain our numerical result which we are 
going to discuss in Section \ref{sec:results} before we 
conclude in Section \ref{sec:conclusions}.

\section{Calculational setup}
\label{sec:setup}
The results presented in this paper have been obtained by combining the two tools
\GoSam~\cite{Cullen:2011ac,Cullen:2014yla} and \Sherpa~\cite{Gleisberg:2008ta}
which allows for a fully automated calculation of cross section and observables and next-to-leading order in QCD as well
as in the electroweak coupling.
\GoSam is a package which generates the code for the numerical evaluation of
the one loop scattering amplitudes starting from the Feynman diagrams,
generated with \QGraf~\cite{Nogueira:1991ex} and further processed with
\FORM~\cite{Vermaseren:2000nd,Kuipers:2012rf} and
\Spinney~\cite{Cullen:2010jv} to perform necessary algebraic
manipulations to obtain an optimized expression for the matrix elements.
For the integrand reduction of the diagrams we use the \Ninja
library~\cite{Peraro:2014cba}, an implementation of the technique of integrand
reduction via Laurent expansion~\cite{Mastrolia:2012bu,vanDeurzen:2013saa}.
Alternatively one can choose other reduction strategies such as OPP reduction
method~\cite{Ossola:2006us,Mastrolia:2008jb,Ossola:2008xq} which is
implemented in $d$ dimensions in \Samurai~\cite{Mastrolia:2010nb}, or methods based on
tensor integral reduction as implemented in
\GolemNF~\cite{Heinrich:2010ax,Binoth:2008uq,Cullen:2011kv,Guillet:2013msa}.
We have used \OneLoop~\cite{vanHameren:2010cp} to evaluate the scalar integrals.

\begin{figure}[t!]
  \begin{tabular}{ccccc}
    \includegraphics[width=0.288\textwidth]{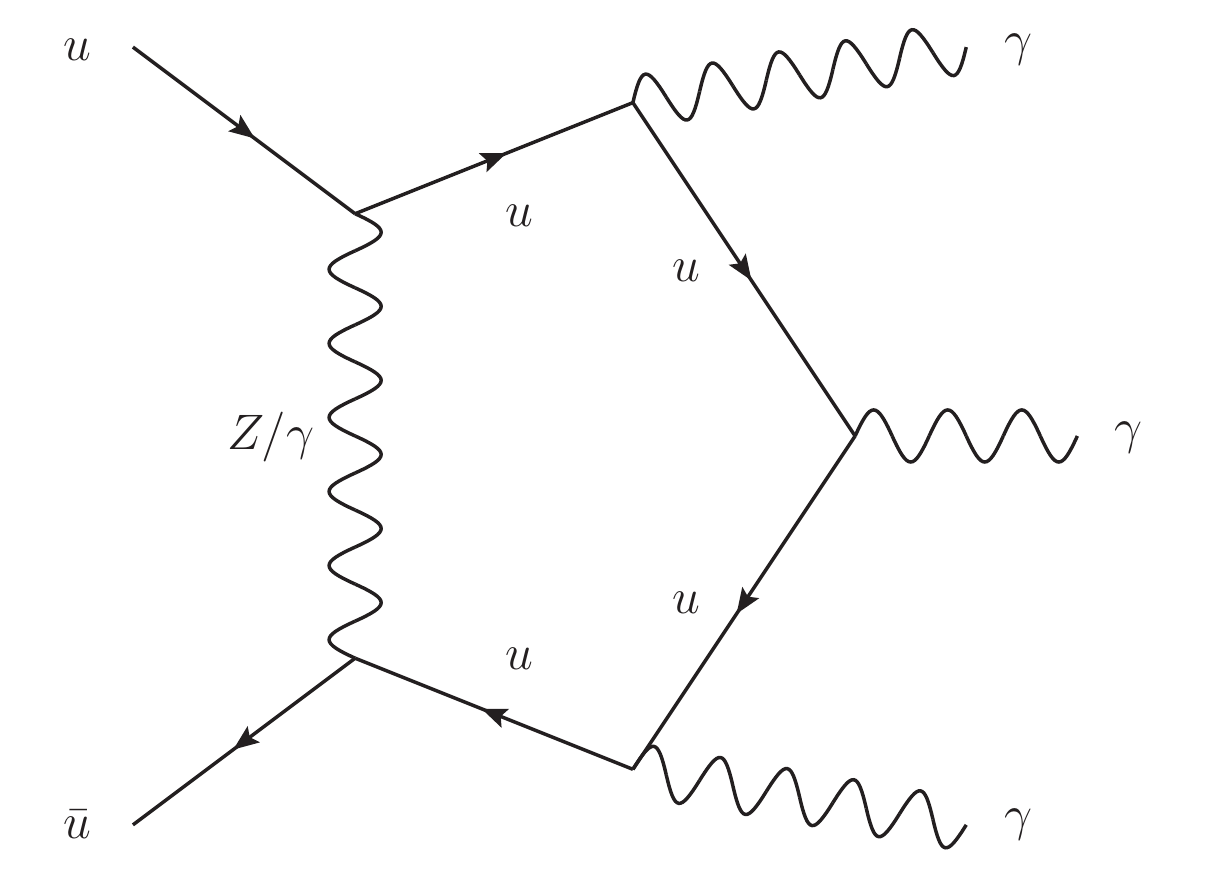} & &
    \includegraphics[width=0.288\textwidth]{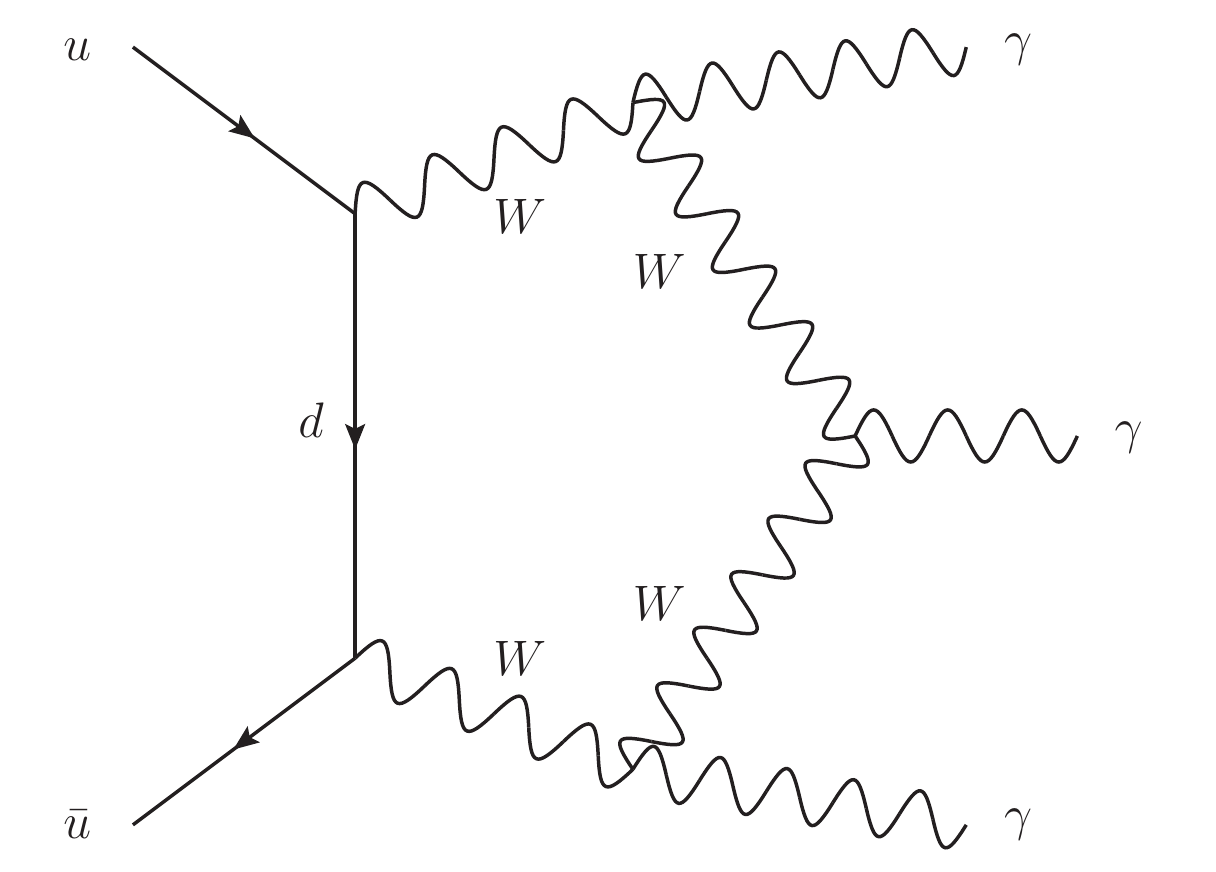} & &
    \includegraphics[width=0.288\textwidth]{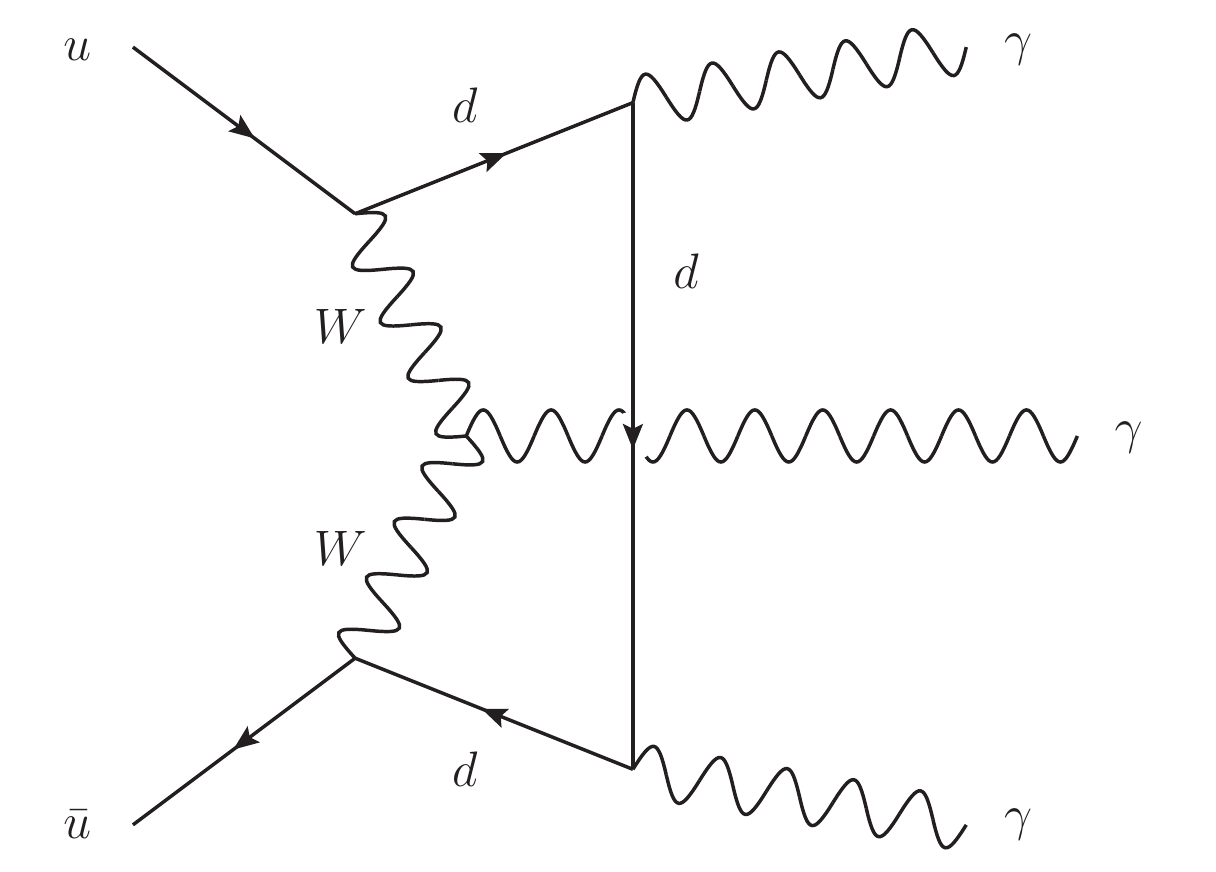} \\
  \end{tabular}
  \caption{
    Sample diagrams of electroweak virtual corrections to triple 
    photon production.
    \label{fig:aaa:amps}
  }
\end{figure}

\begin{figure}[t!]
  \begin{tabular}{ccccc}
    \includegraphics[width=0.288\textwidth]{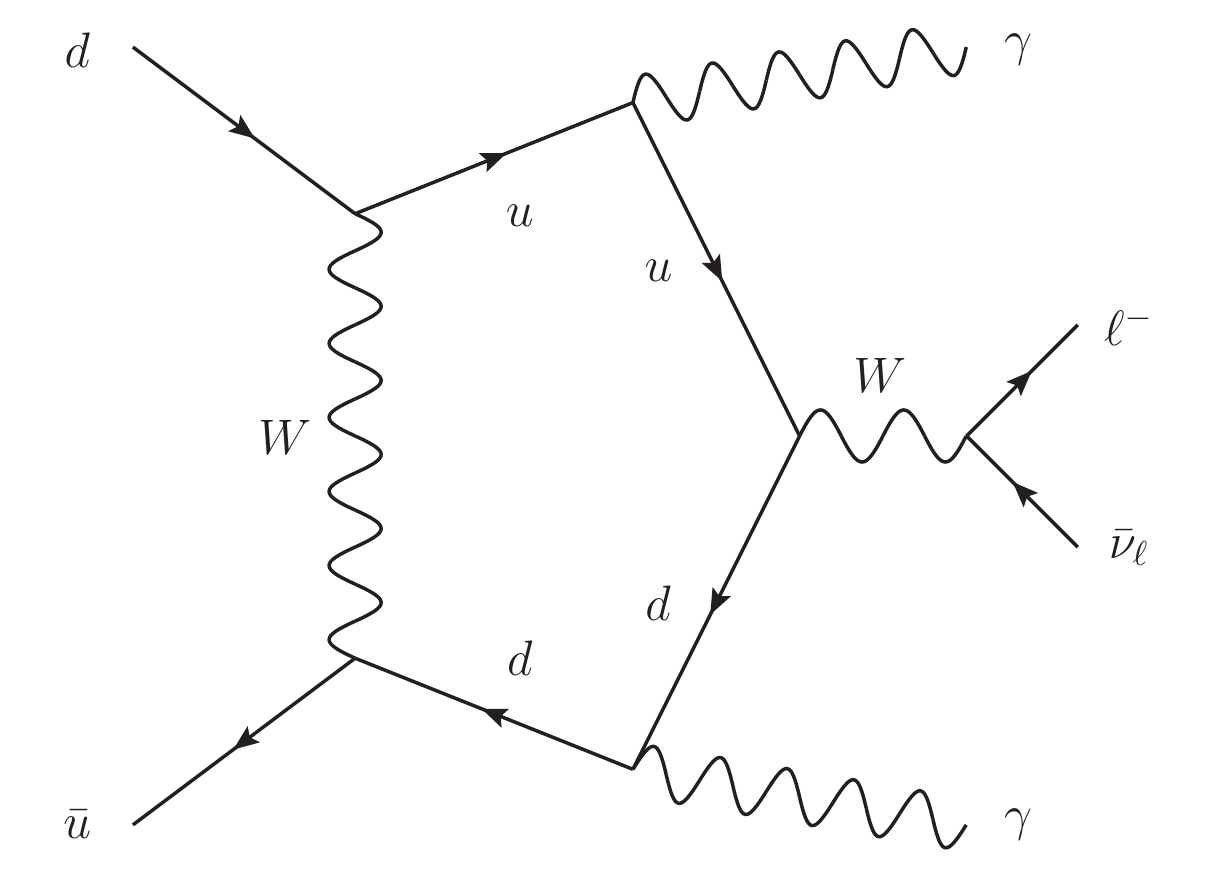} & &
    \includegraphics[width=0.288\textwidth]{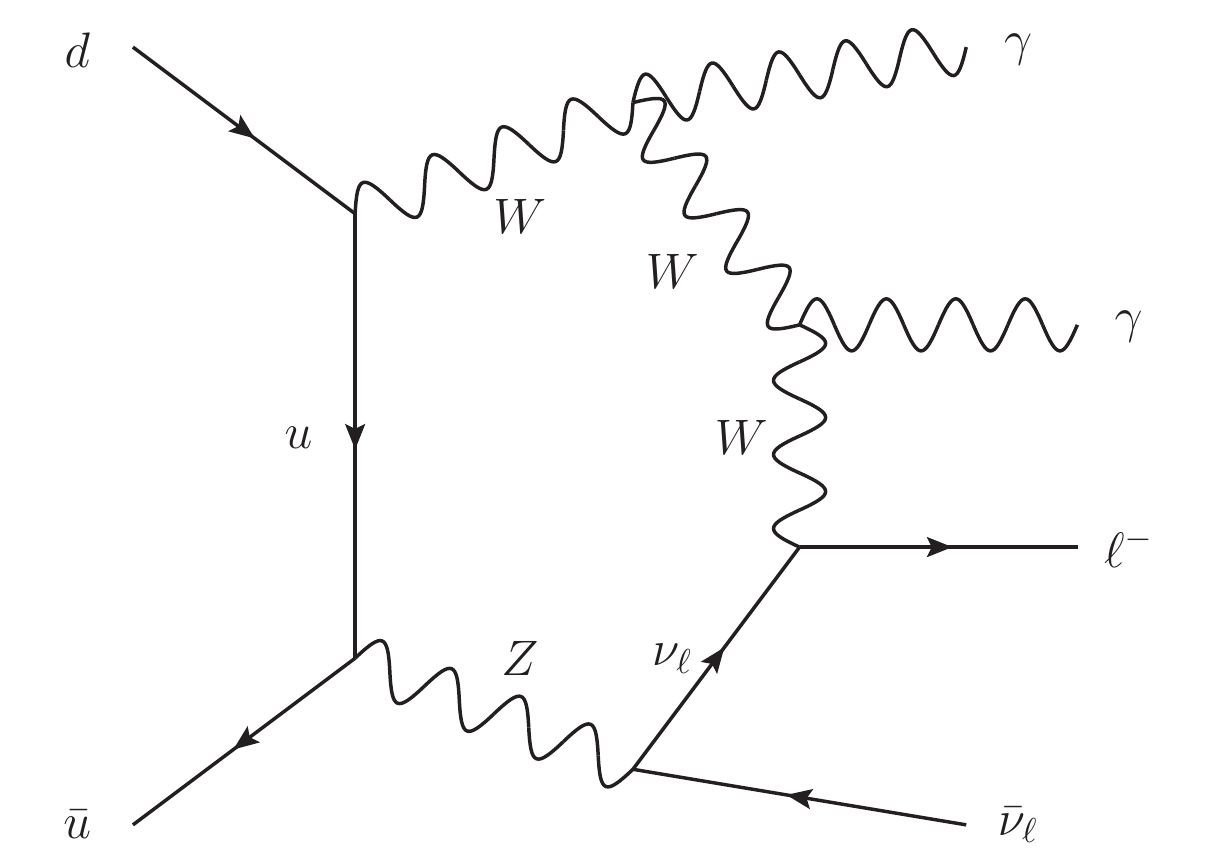} & &
    \includegraphics[width=0.288\textwidth]{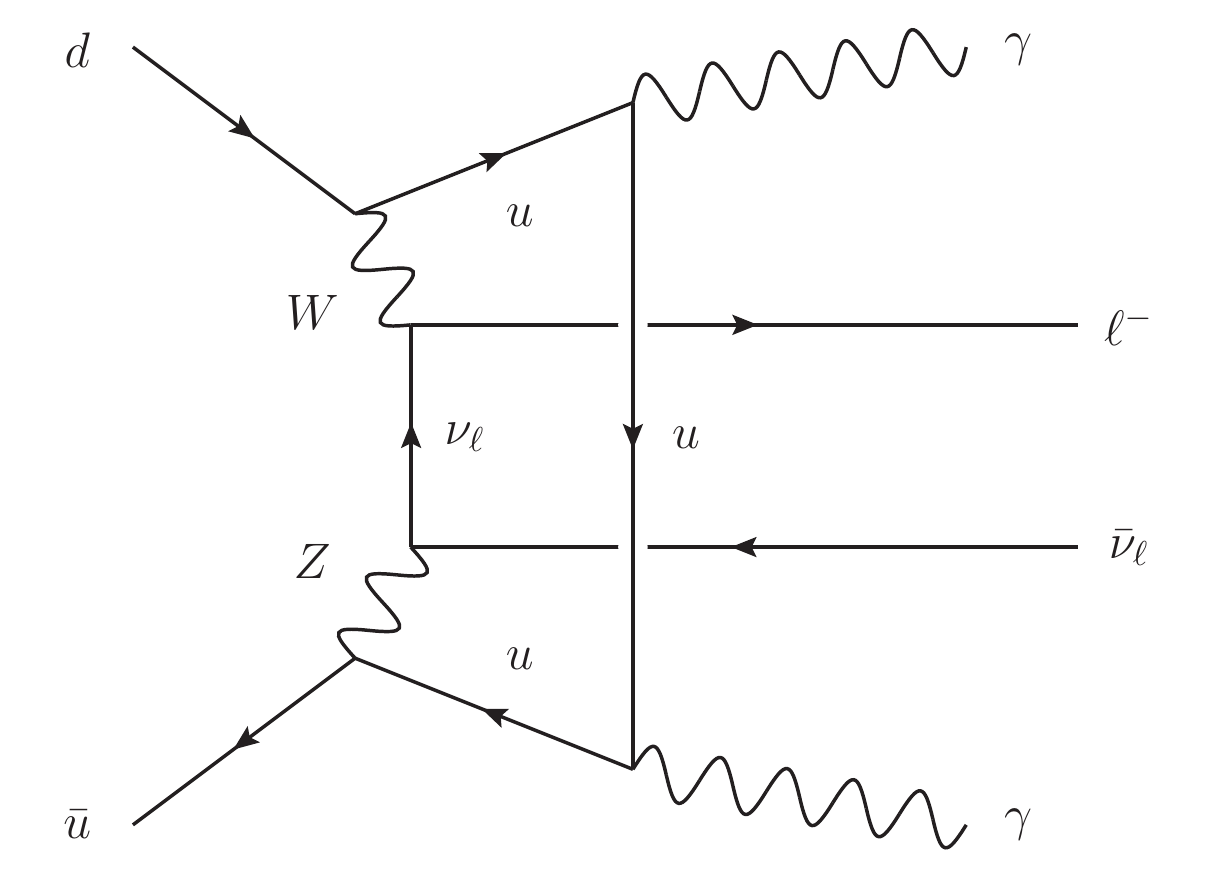} \\
  \end{tabular}
  \caption{
    Sample diagrams of electroweak virtual corrections to diphoton 
    production in association with a lepton-neutrino pair.
    \label{fig:aaw:amps}
  }
\end{figure}

\begin{figure}[t!]
  \begin{tabular}{ccccc}
    \includegraphics[width=0.288\textwidth]{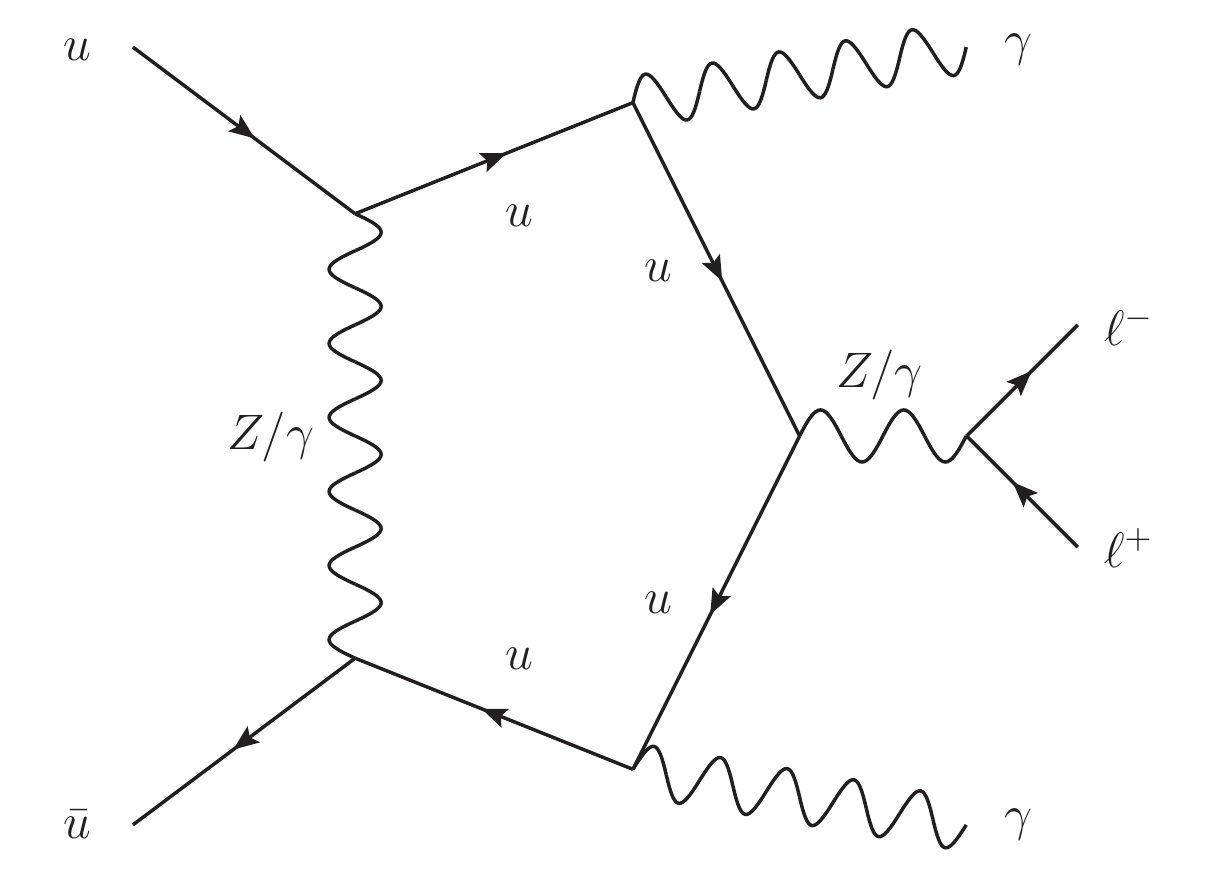} & &
    \includegraphics[width=0.288\textwidth]{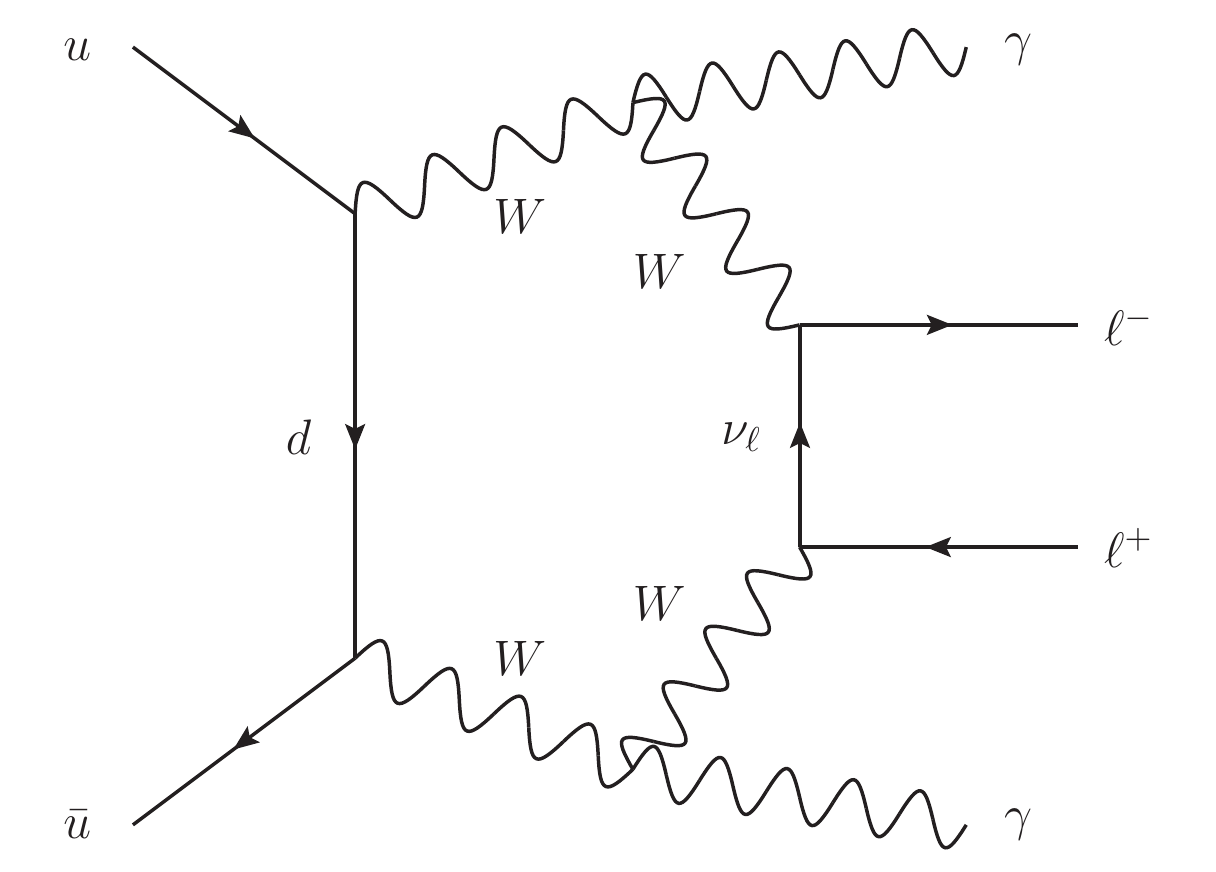} & &
    \includegraphics[width=0.288\textwidth]{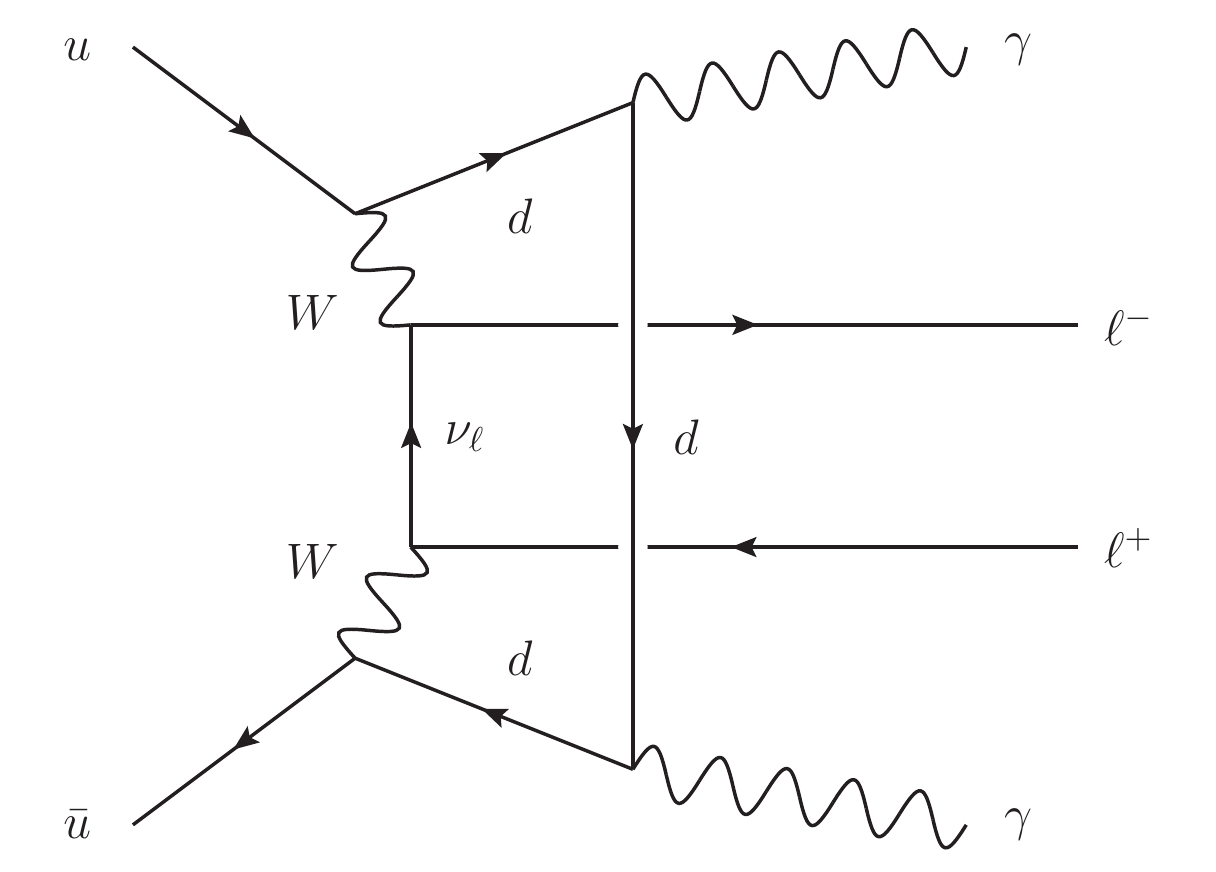} \\
  \end{tabular}
  \caption{
    Sample diagrams of electroweak virtual corrections to diphoton 
    production in association with a lepton-pair.
    \label{fig:aaz:amps}
  }
\end{figure}

A selection of one-loop amplitudes contributing to the three 
considered processes are shown in Figures 
\ref{fig:aaa:amps}--\ref{fig:aaz:amps}. 
The highest point loop integrals occurring in $\gamma\gamma\gamma$ 
production are pentagons, while both $\gamma\gamma e^-\bar\nu_e$ 
and $\gamma\gamma e^+e^-$ productions include up to hexagons at 
NLO EW. 

\Sherpa, on the other hand, provides the tree-level matrix 
elements, infrared subtraction, process management and phase-space 
integration of all contributions to all processes considered in 
this publication through its tree-level matrix element generator 
\textsc{Amegic} \cite{Krauss:2001iv}. 
Its inbuilt infrared subtraction is performed in the QED 
generalisation of the Catani-Seymour scheme~\cite{Catani:1996vz,
  Dittmaier:1999mb,Gleisberg:2007md,Kallweit:2014xda,
  Kallweit:2015dum,Kallweit:2017khh,Schonherr:2017xxx}
and includes the appropriate initial state mass factorisation 
counter terms.
Both programs, \Sherpa and \GoSam, are interfaced through a 
dedicated interface based on the 
Binoth Les Houches Accord~\cite{Binoth:2010xt,Alioli:2013nda}. 
Cross-checks of the tree-level matrix elements of \GoSam and 
\Sherpa and the renormalized pole coefficients of the virtual 
corrections of \GoSam and the infrared poles of \Sherpa have 
been performed for several phase space points spanning 
multiple kinematic regimes and we have found excellent 
agreement.

This combination of tools was previously used to calculate the 
NLO QCD corrections to $ZZ+\text{jet}$, $t\bar{t}+0,1\,\text{jets}$, 
$W^{+}W^{-}b\bar{b}$ and $h+0,1,2,3\,\text{jets}$ production 
in \cite{Binoth:2009wk,Hoeche:2013mua,Heinrich:2013qaa,
  Greiner:2015jha,Greiner:2016awe,Heinrich:2017bqp} and the 
NLO EW corrections to $\gamma\gamma+0,1,2\,\text{jets}$ production 
in \cite{Chiesa:2017gqx}.

\section{Results}
\label{sec:results}

In this section we present numerical results for the NLO QCD and NLO EW 
corrections to all three production processes of a diphoton pair in 
association with a third vector boson, a third photon, a $W$ or a $Z$ 
boson, at the LHC at a centre-of-mass energy of 13\,TeV. 
In case of an accompanying $W$ or a $Z$ boson, we consider the full 
off-shell leptonic final state, i.e.\ lepton-neutrino or lepton-pair 
production.
All results are obtained in the Standard Model using the complex-mass 
scheme \cite{Denner:2005fg} with the following input parameters
\begin{center}
  \begin{tabular}{rclrcl}
    $\alphazero$ &\shortequal& $1/137.03599976$  \qquad &&& \\
    $\Gmu$ &\shortequal& $1.1663787\times 10^{-5}\; \text{GeV}^2$ &&& \\
    $m_W$ &\shortequal& $80.385\; \text{GeV}$       & $\Gamma_W$ &\shortequal& $2.085\; \text{GeV}$ \\
    $m_Z$ &\shortequal& $91.1876\; \text{GeV}$      & $\Gamma_Z$ &\shortequal& $2.4952\; \text{GeV}$ \\
    $m_h$ &\shortequal& $125.0\; \text{GeV}$        & $\Gamma_h$ &\shortequal& $0$\\
    $m_t$ &\shortequal& $173.2\; \text{GeV}$        & $\Gamma_t$ &\shortequal& $0$\;.
  \end{tabular}
\end{center}
While we calculate triple photon production in the \alphazero-scheme, 
we use a mixed scheme for \aaw\ and \aaz production: at LO two 
powers of $\alpha$ are taken in the \alphazero-scheme, while 
one power of $\alpha$ is taken in the \Gmu-scheme. 
The additional power of $\alpha$ in the NLO EW correction 
is evaluated in \alphazero-scheme again.
The virtual amplitudes are renormalised correspondingly.
In all cases both the width of the top quark and the Higgs boson 
can safely be neglected as there are no diagrams containing either 
as $s$-channel propagators which can potentially go on-shell. 
All other lepton and parton masses and widths are set to zero, 
i.e.\ we are working in the five-flavour scheme.
We use the \textsc{CT14nlo} PDF set with $\alphas(m_Z)=0.118$ 
\cite{Dulat:2015mca}, interfaced through LHAPDF6 \cite{Buckley:2014ana}. 
The use of a QCD-only PDF is justified by the fact that, 
at LO, the photon induced corrections are either non-existent 
(\aaa, \aaw) or negligible (\aaz).
This finding will be detailed in Section \ref{sec:results:aaz}.

We define our central scales through
\begin{equation}
  \label{eq:murfdef}
    \muR^0 = \muF^0 = 
      \tfrac{1}{2}\,H_T' \;. 
\end{equation}
In the case of the triple photon process $ H_\mathrm{T}' $ is just given by the scalar sum of all final state 
transverse momenta, for the two processes with the massive vector bosons it is defined as 
\begin{equation}
  \label{ew:defHT}
  \begin{split}
    H_\mathrm{T}' = E_\mathrm{T}^V + \sum_{\gamma,q,g} p_{\mathrm{T},i}
  \end{split}
\end{equation}
with $\left.E_\mathrm{T}^W\right.^2=(p_\ell+p_\nu)^2$ and 
$\left.E_\mathrm{T}^Z\right.^2=(p_{\ell^+}+p_{\ell^-})^2$ 
in full analogy to the case of the case of vector boson production 
in association with jets \cite{Berger:2009ep}.
As the Born process in each case has no $\muR$ dependence, we 
do not expect the choice of scale to have a significant influence 
on the size of the relative QCD and EW corrections.
We calculate the leading order cross section $\rd\sigma_\text{LO}(\muF)$, 
which only depends on the factorisation scale $\muF$, the 
NLO QCD differential correction factor $\deltaQCD(\muR,\muF)$, 
introducing the additional $\muR$-dependence, and the NLO EW 
differential correction factor $\deltaEW(\muF^0)$. 
To estimate the impact of yet-to-be-calculated higher-order 
corrections we vary the free scales $\muR$ and $\muF$ 
by the conventional factor of two around their central values 
$\muR^0$ and $\muF^0$, respectively.
We do not vary the factorisation scale for the determination 
of $\deltaEW$ as the inherent, albeit normally phenomenologically 
irrelevant, stabilisation of the $\muF$-dependence at NLO EW 
is not reflected in our chosen PDF.
Hence, our NLO EW result exhibits the exact same $\muF$-dependence 
as the LO result.
We define 
\begin{equation}
  \label{eq:defnlo}
  \begin{split}
    \rd\sigma_\text{NLO QCD}
    \,=\;&\rd\sigma_\text{LO}\left(1+\deltaQCD\right)\\
    \rd\sigma_\text{NLO EW}
    \,=\;&\rd\sigma_\text{LO}\left(1+\deltaEW\right)\\
    \rd\sigma_\text{NLO \QCDpEW}
    \,=\;&\rd\sigma_\text{LO}\left(1+\deltaQCD+\deltaEW\right)\\
    \rd\sigma_\text{NLO \QCDtEW}
    \,=\;&\rd\sigma_\text{LO}\left(1+\deltaQCD\right)\left(1+\deltaEW\right)\;.
  \end{split}
\end{equation}
Therein, the difference between NLO \QCDpEW\ and NLO \QCDtEW, which 
is of relative $\order(\alphas\alpha)$, can serve as an indicator 
of the potential size of unknown corrections at that order.

In Table \ref{tab:xsec} we quote the inclusive cross sections for all 
three processes. 
The set of fiducial cuts for each process is detailed in its 
respective subsection below. 
We note that the NLO QCD corrections for both triple photon 
production and \aaw\ production are strongly jet veto dependent, 
a result that was previously discussed in great detail in 
\cite{Bozzi:2011en,Bozzi:2011wwa}, and will be revisited in the 
following. 
A much milder jet veto dependence is found for \aaz\ production. 
The electroweak corrections to inclusive cross sections are 
generally much smaller, ranging from $0.6\%$ (\aaa) to $-1.8\%$ (\aaw) 
and $-4.4\%$ (\aaz).

\begin{table}[t!]
  \centering
\begin{tabular}{l||c|c|c}
  & $\;\;pp \to \gamma \gamma\gamma\;\;$
  & $\;\;pp \to \gamma \gamma e^-\bar\nu_e\;\;$
  & $\;\;pp \to \gamma \gamma e^+e^-\;\;$ \\
  \hline\hline
  $\sigma_\text{LO}\;\;[\text{fb}]\vP$ & $5.56_{-0.36}^{+0.30}$ & $0.92_{-0.07}^{+0.06}$ & $4.21_{-0.41}^{+0.36}$ \\
  \hline\hline
  $\delta_\text{QCD}\;\;[\%]\;\;\pTveto=\infty\vP$ & $139_{-27}^{+24}$ & $111_{-24}^{+21}$ & $27_{-18}^{+13}$ \\
  \hline
  $\delta_\text{QCD}\;\;[\%]\;\;\pTveto=30\,\text{GeV}\vP$ & $\hspace*{\unitcharwidth}35_{-13}^{+\hspace*{\unitsuperscriptwidth}7}$ & $\hspace*{\unitcharwidth}41_{-14}^{+\hspace*{\unitsuperscriptwidth}8}$ & $19_{-17}^{+11}$ \\
  \hline
  $\delta_\text{EW}\;\;[\%]\vP$ & $0.6$ & $-1.8$ & $-4.4$ \\
\end{tabular}
  \caption{
    Total cross sections at LO, NLO QCD and NLO EW for $\gamma\gamma\gamma$, 
    $\gamma\gamma e^-\bar\nu_e$ and $\gamma\gamma e^+e^-$
    production at 13\,TeV at the LHC.
    \label{tab:xsec}
  } 
\end{table}

\subsection[\texorpdfstring{$\gamma\gamma\gamma$}{aaa} production]
           {$\boldsymbol{\gamma\gamma\gamma}$ production}
\label{sec:results:aaa}

The triple photon production process is defined 
by the presence of three identified photons in the 
central detector. 
To this end we use the smooth cone isolation 
criterion \cite{Frixione:1998jh}, limiting the amount of 
hadronic activity in a cone $R_\gamma$ to
\begin{equation}
  \begin{split}
    E_{{\rm had, max}} (r_{\gamma})
    = \epsilon\, \pT^{\gamma} \left( \frac{1-\cos r_\gamma}
				       {1-\cos R_\gamma}\right)^{n}\;,
  \end{split}
  \label{eq:frix}
\end{equation}
where $r_{\gamma}$ denotes the angular separation between the photon and 
the parton, with 
\begin{equation}
  \label{eq:coneparams}
  \begin{split}
    R_{\gamma}=0.4\;, \quad \epsilon = 0.05\;, \quad n = 1\;, 
  \end{split}
\end{equation}
to define isolated photon candidates.
These candidates are then ordered in transverse momentum. 
We require at least three such candidates within $|\eta|<2.37$, 
the leading one of which needs $\pT>40\,\text{GeV}$, 
while the all subleading ones need only 
$\pT>30\,\text{GeV}$. 
Finally, a pairwise separation of $\Delta R(\gamma_i,\gamma_j)>0.4$ 
between all identified photons is required.
It is worth noting that at NLO EW it is possible to find more 
than three isolated photons, in which case any combination may 
fulfill the above criteria. 

\begin{figure}[t!]
  \centering
  \includegraphics[width=0.32\textwidth]{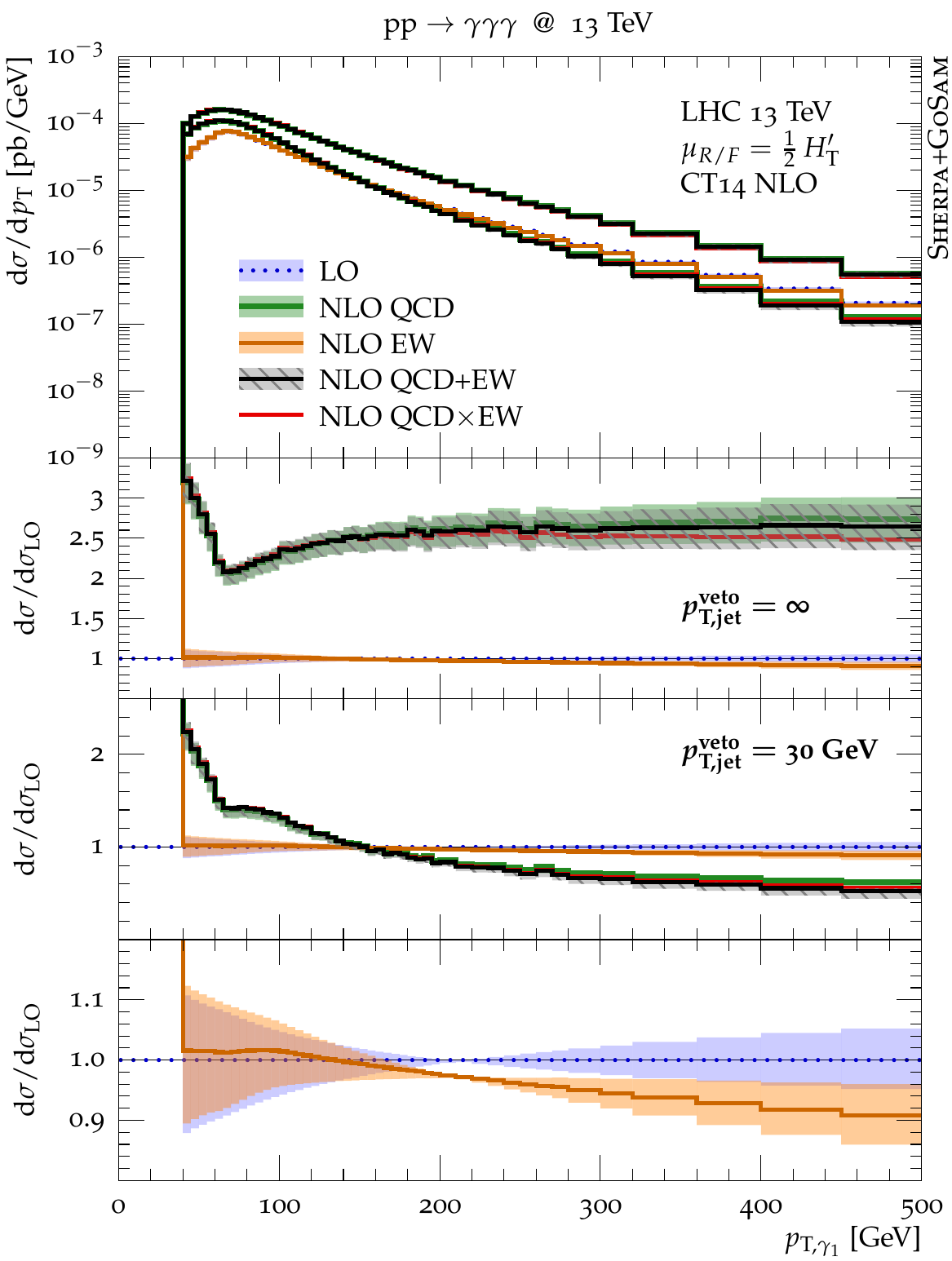}
  \includegraphics[width=0.32\textwidth]{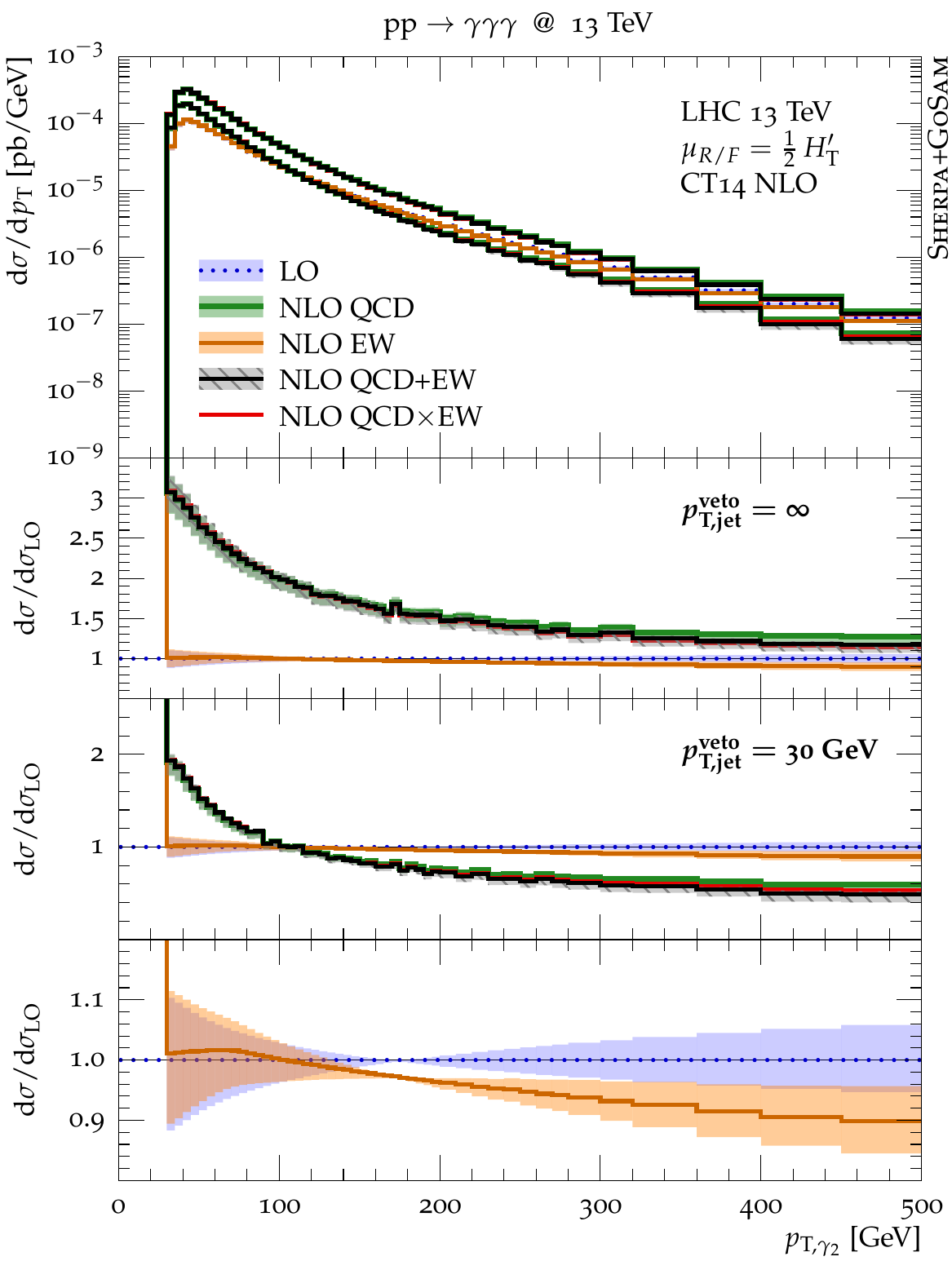}
  \includegraphics[width=0.32\textwidth]{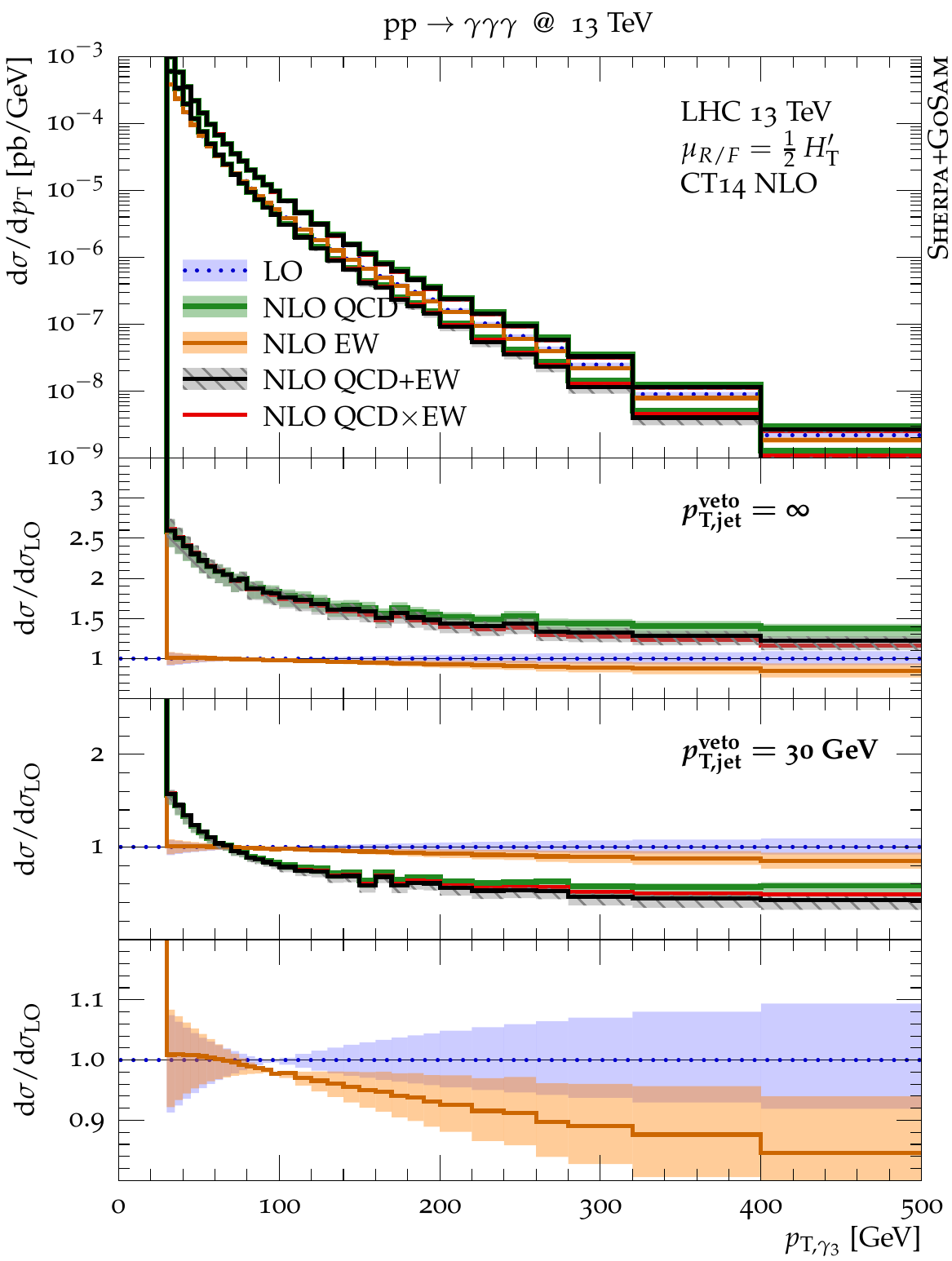}
  \caption{
    Transverse momentum of the leading (left), subleading (centre) 
    and third leading (right) photon 
    in triple photon production at the LHC at 13\,TeV. 
    The distributions are shown at LO (blue), NLO QCD (green), 
    NLO EW (orange), NLO \QCDpEW\ (black) and NLO \QCDtEW\ (red) 
    including scale uncertainties. The top panel displays the 
    absolute predictions for both the case with and without the 
    application of a jet veto of $p_\mathrm{T,jet}^\mathrm{veto}=\text{30\,GeV}$, 
    the latter ones are easily recognisable 
    due to their much reduced rate. The top ratio plot details 
    the relative corrections to the leading order cross section 
    without applying the jet veto, while the centre ratio plot 
    applies the jet veto. 
    The lower ratio highlights the size of the electroweak corrections.
    \label{fig:aaa:pt}
  }
\end{figure}

Figure \ref{fig:aaa:pt} displays the transverse momenta of the 
first three leading photons. 
The first observation is that, with the chosen set of cuts, 
the leading and subleading photon are of similar hardness, 
and generally much harder than the third leading photon. 
This suggest that we select primarily diphoton production events 
which is accompanied by a third, mostly bremsstrahlung, photon. 
The NLO QCD corrections exhibit a handful of interesting features. 
In the absence of a jet veto the fixed-order calculation 
exhibits huge correction factors, mainly induced by the 
opening of new channels in the real emission. 
Additionally, kinematic constraints present at LO \footnote{
  At leading order the leading photon needs to be in a different 
  hemisphere than both the subleading and third leading photon. 
  Thus, $\Delta\phi_{\gamma_1\gamma_2}$ and $\Delta\phi_{\gamma_1\gamma_3}$ 
  must be larger that $\tfrac{1}{2}\,\pi$.
} are released and 
lead to a larger phase space that can be populated. 
These findings mandate the inclusion of at least the $\aaa+\text{jet}$ 
production process at NLO QCD to arrive at a reliable description 
of inclusive \aaa\ production, and thus either a NNLO QCD calculation 
or a multijet merging ansatz \cite{Hoeche:2012yf,Kallweit:2015dum}. 
While the inclusive QCD corrections at very small transverse momenta 
are universally large for all three leading photon \pT-spectra 
($\deltaQCD\approx 2$), they remain at exorbitantly large 
($\deltaQCD\approx 1.5$) throughout the considered range 
only for the leading jet \pT. 
For both subleading photons the QCD corrections are quickly 
decreasing, leveling out at a approximately $20\%$ at large transverse 
momenta.
In the presence of a restrictive jet veto the very low transverse 
momentum region still experiences large correction of about 
$\deltaQCD\approx 1$. 
As transverse momenta are increasing, however, the QCD corrections now turn 
negative reaching now $-50-60\%$ and for all three photons. 

The electroweak corrections, due to the absence of the opening 
of large new channels at the next-to-leading order, are dominated 
by the virtual corrections. 
Consequently, the release of the LO phase space restrictions 
in the real emission corrections only plays a minor role.
For all three photons the electroweak corrections are small but 
positive at small transverse momenta and exhibit the usual 
Sudakov shape at large transverse momenta.
They reach $-10\%$ for the leading and subleading photon and 
$-20\%$ for the third photon at $\pT=500\,\text{GeV}$. 
More importantly, beyond $\pT\gtrsim 170$, 140 and 80\,GeV 
for the first, second and third leading photon, respectively, 
the electroweak corrections are not covered by the LO 
uncertainty estimate.
Due to the different sizes of the QCD and electroweak 
corrections, the additive and multiplicative combination 
of corrections lead to very similar results.

\begin{figure}[t!]
  \centering
  \includegraphics[width=0.32\textwidth]{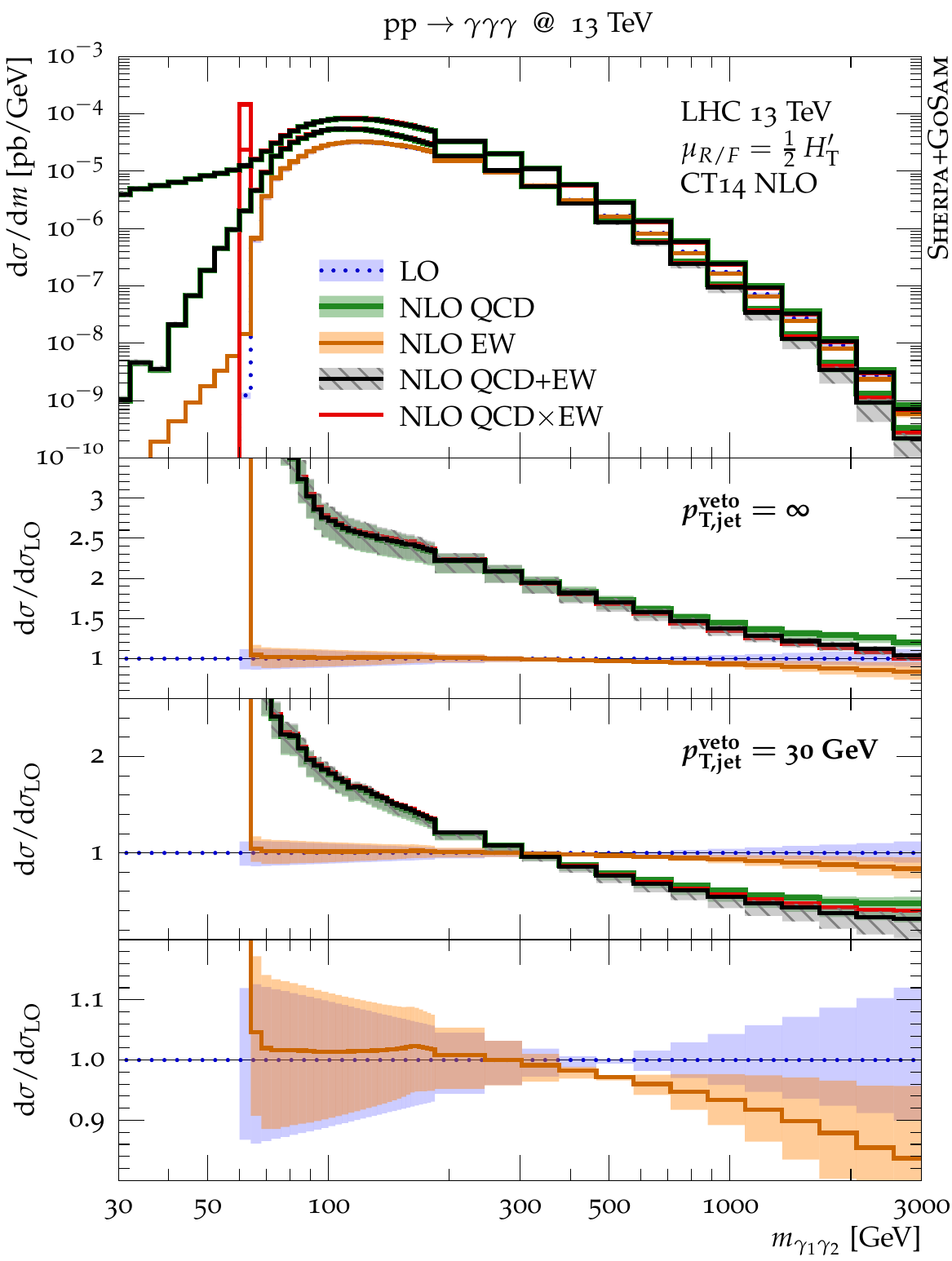}
  \includegraphics[width=0.32\textwidth]{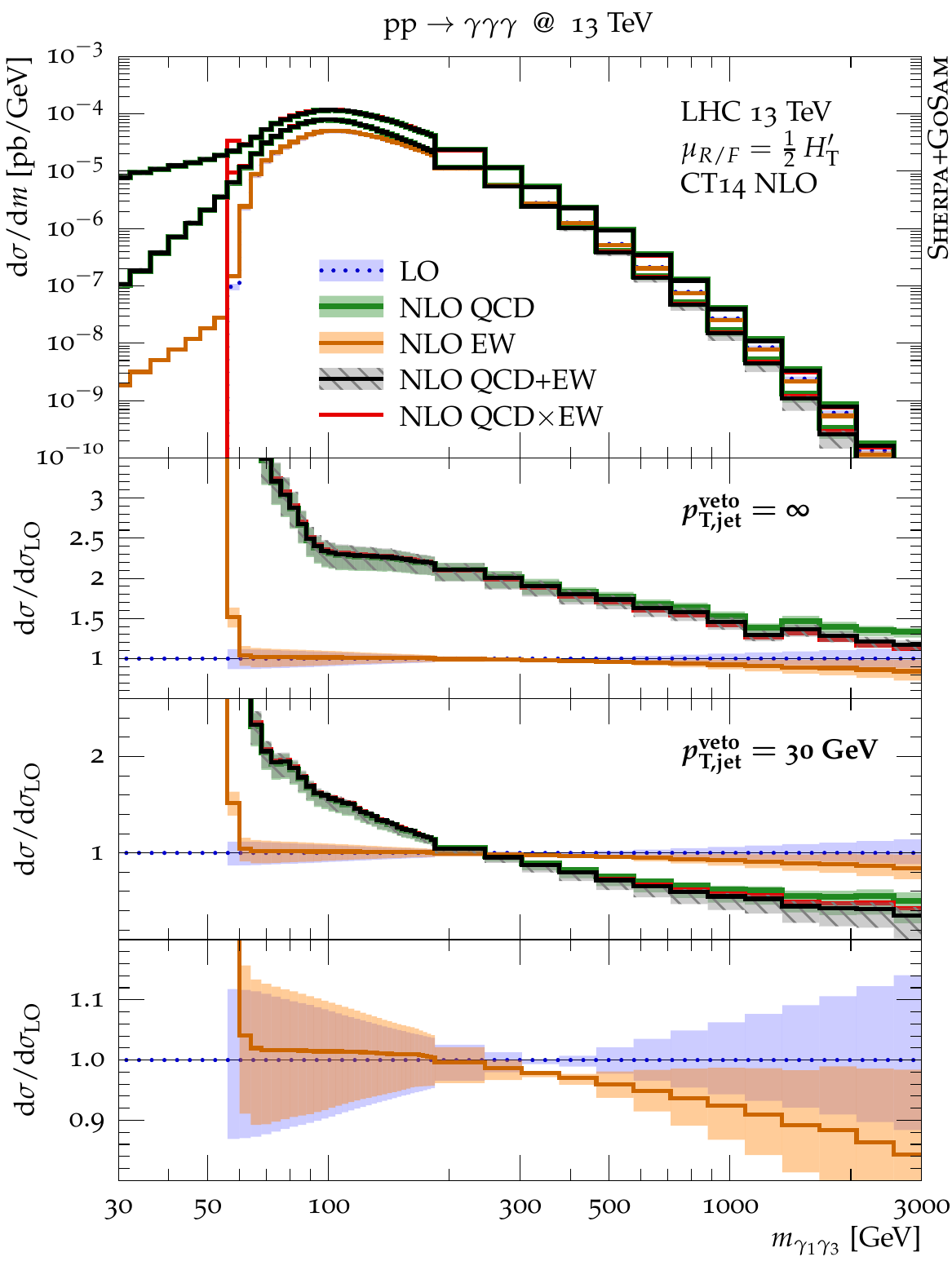}
  \includegraphics[width=0.32\textwidth]{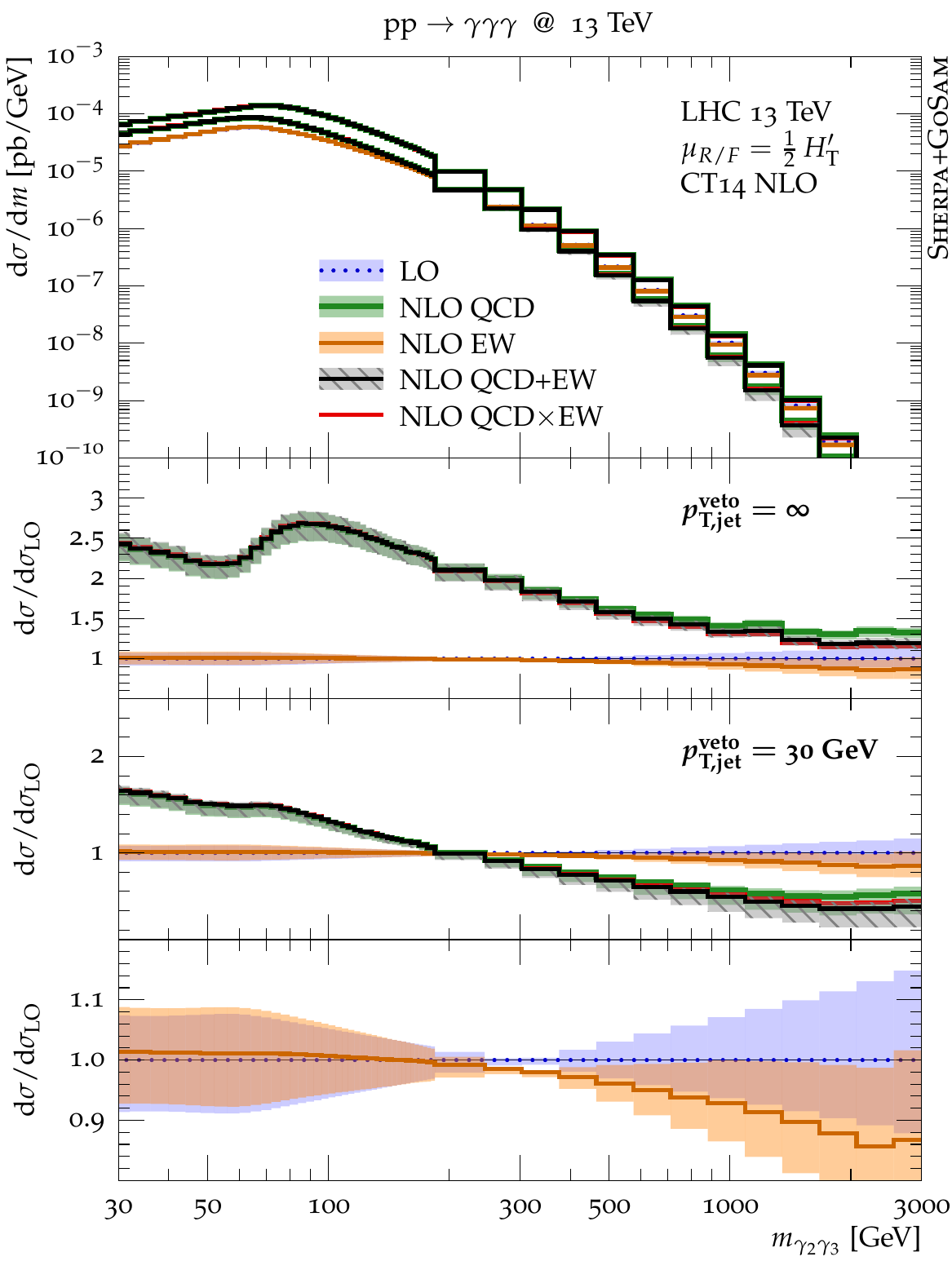}
  \caption{
    Pairwise invariant mass of the leading and subleading photon (left),
    leading and third leading photon (centre), subleading and third leading 
    photon (right) 
    in triple photon production at the LHC at 13\,TeV. 
    Details as in Fig.\ \ref{fig:aaa:pt}.
    \label{fig:aaa:myy}
  }
\end{figure}

Figure \ref{fig:aaa:myy} continues with the three combinations of 
diphoton invariant masses. 
The \pT\ and $\Delta R$ requirements 
of the event selection induce a minimum in the distributions 
at leading order. 
The region below can only be filled if a fourth particle is 
present, as is the case in both the QCD and electroweak real 
emission corrections, leading to simultaneously huge 
corrections \deltaQCD\ and \deltaEW\ as the Born cross 
section vanishes. 
Due to this behaviour, the multiplicative combination of 
corrections, NLO \QCDtEW, ceases to be well defined and 
spikes in the distribution are visible.
The distributions below, where $\rd\sigma_\text{LO}=0$, 
are ill-defined. 
In consequence, for distributions where kinematic boundaries 
exist at leading order, but are lifted at higher orders, 
the multiplicative combination does not present a viable 
option for describing the observable 
throughout phase space. 

The QCD corrections themselves again exceed $200\%$ at 
small invariant masses, already well before the above 
described kinematic boundary effect takes hold. 
As the invariant masses are increasing, the 
QCD corrections are dropping to $20-30\%$ for the inclusive 
selection. 
The structure the QCD corrections exhibit around 80-90\,GeV 
in all three diphoton-pair invariant masses are induced by 
the acceptance cuts.
In the presence of the jet veto, the QCD corrections are 
reduced and turn negative beyond $m_{\gamma\gamma}\gtrsim 300\,\text{GeV}$ 
reaching around $-40\%$ at 1\,TeV. 
The electroweak corrections, on the other hand, are 
again moderate, ranging from $+1.5\%$ between 70 and 200\,GeV 
for $m_{\gamma_1\gamma_2}$ and $m_{\gamma_1\gamma_3}$ and 
0 and 100\,GeV for $m_{\gamma_2\gamma_3}$. 
$m_{\gamma_1\gamma_2}$ exhibits a small rise in the correction 
at $2m_W$ due to resonant box diagrams in that region. 
This feature is also present in the electroweak corrections 
to diphoton production in this observable \cite{Chiesa:2017gqx}. 
At large transverse momentum the usual Sudakov logarithms 
are recovered, resulting in corrections of around 
$-8\%$ at 1\,TeV for all photon-pair invariant masses. 
Most importantly, beyond $m_{\gamma\gamma}\gtrsim 300\,\text{GeV}$, 
the electroweak corrections are again not covered by the LO 
uncertainty estimate.

\begin{figure}[t!]
  \centering
  \includegraphics[width=0.32\textwidth]{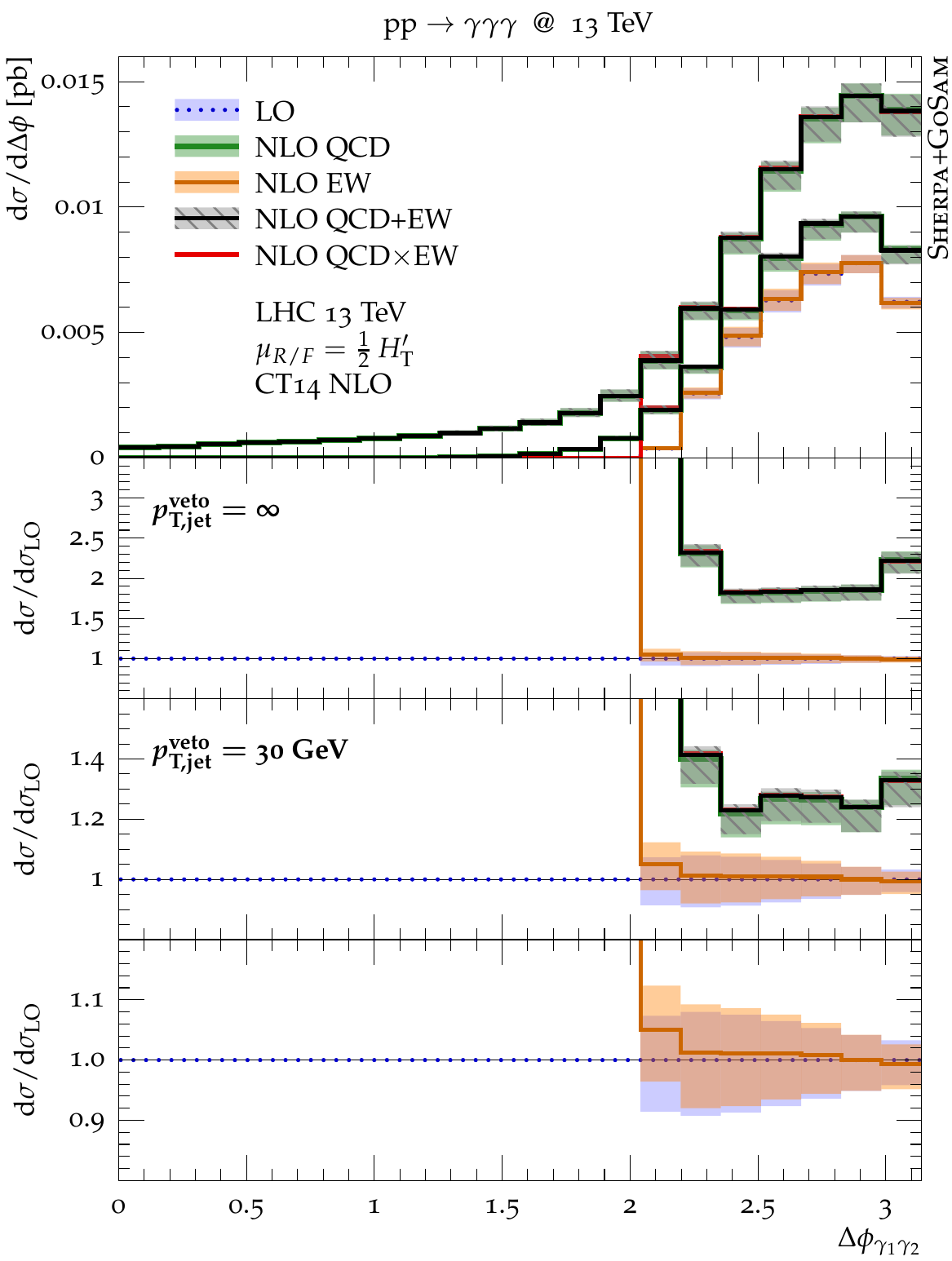}
  \includegraphics[width=0.32\textwidth]{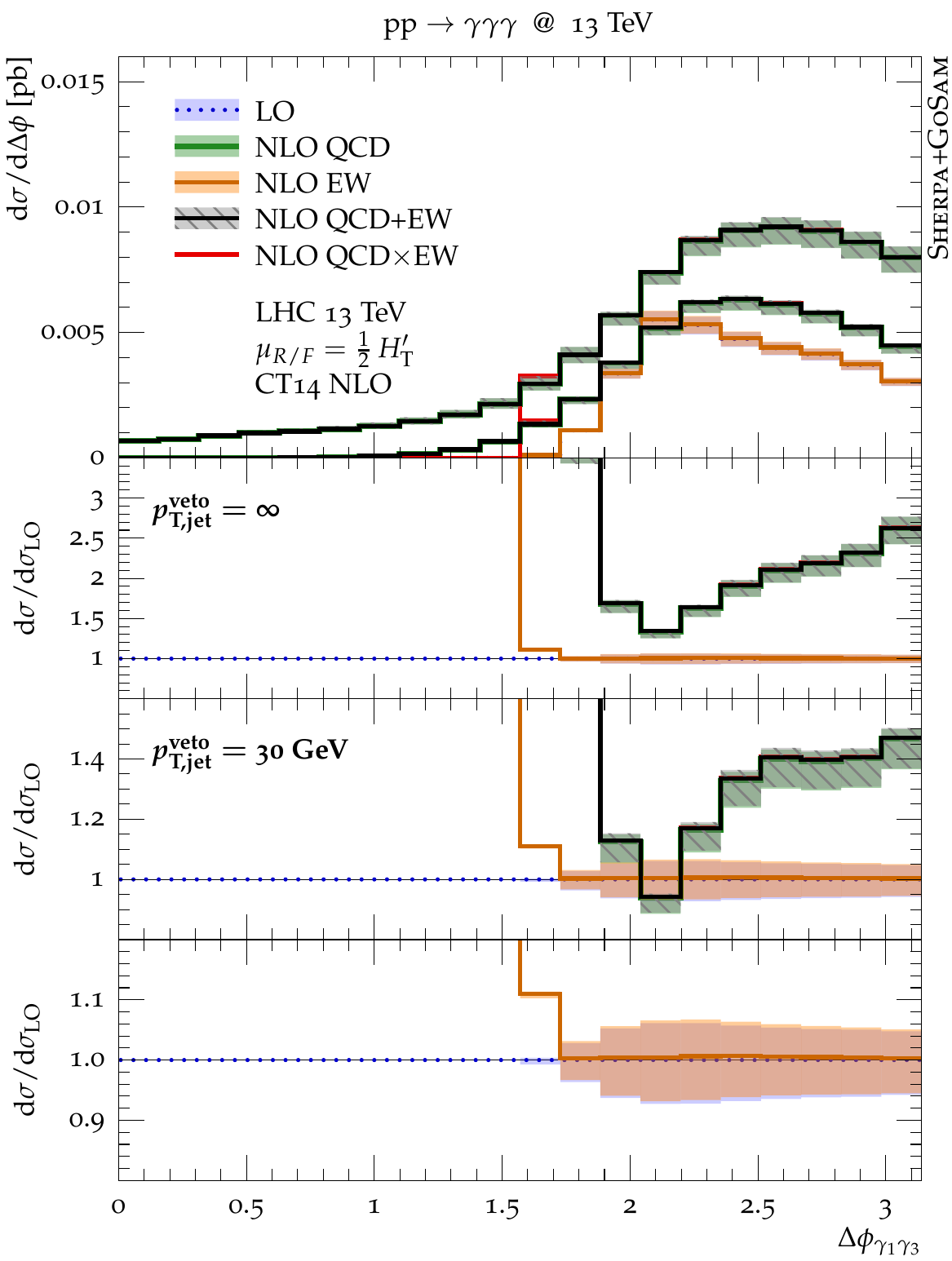}
  \includegraphics[width=0.32\textwidth]{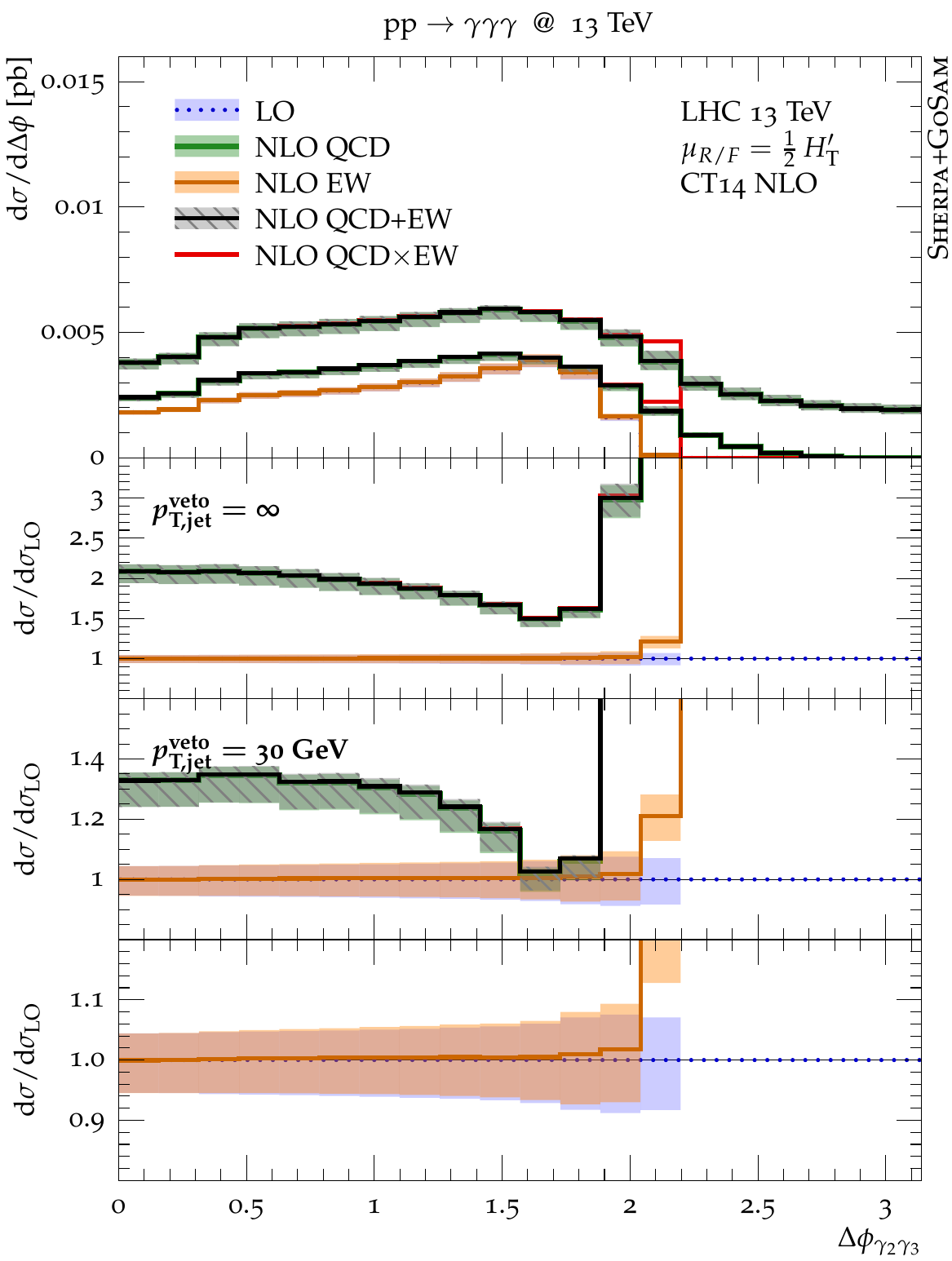}
  \caption{
    Azimuthal separation of the leading and subleading photon (left),
    leading and third leading photon (centre), subleading and third leading 
    photon (right) 
    in triple photon production at the LHC at 13\,TeV. 
    Details as in Fig.\ \ref{fig:aaa:pt}.
    \label{fig:aaa:dphi}
  }
\end{figure}

Finally, in Figure \ref{fig:aaa:dphi} we show the azimuthal 
separation $\Delta\phi$ between all three diphoton pairs. 
Similar features as before are visible as both the NLO QCD 
and NLO EW corrections relax the kinematic boundaries of 
the leading order calculation. 
Especially the azimuthal separation of the leading and 
third leading photon receives substantial shape corrections 
throughout the entire spectrum, with and without the 
presence of a jet veto. The angular distribution between leading photon
and the second or third subleading photon exhibit a kinematical edge at $\pi/2$ which 
is relaxed at NLO. 
This kinematical edge can be understood in the following way:
 Let us consider the first observable, the angular separation between 
the leading and the subleading photon, the argument also holds for the separation between
the leading and the third leading photon. The third leading photon has to recoil against the
system of the two leading photons. Let us go into the limit where the three photons have
a very similar transverse momentum and consider the configuration where the leading
photon is back-to-back to the third leading photon. Then the second photon must 
be perpendicular to this axis. If the third photon has less transverse momentum, it 
needs the second photon to recoil against the leading photon. Therefore it gets closer
to the third photon and further away from the leading photon. $\pi/2$ is therefore the
minimal distance the second photon can have to the leading one.  Only at NLO where
one can have additional radiation this constraint is relaxed. We will see later in Fig.
\ref{fig:aaz:dphi} that this situation is also present when one replaces one of the subleading
photons by a $Z$ boson. 
The electroweak corrections are negligible for this 
observable.

\subsection[\texorpdfstring{$\gamma\gamma\ell\nu$}{aalnu} production]
           {$\boldsymbol{\gamma\gamma\ell\nu}$ production}
\label{sec:results:aaw}

Next we move on to diphoton production in association with 
a $W$ boson decaying leptonically.
In this context, we will representatively focus on $W^-$ bosons 
as we do not expect qualitatively different results for $W^+$ 
bosons. 
We define our fiducial phase space by following experimental setups \cite{Aad:2015uqa}. 
First, we require the presence of exactly one charged lepton, 
dressed with all photons in a cone of size $R=0.1$, with 
$\pT>20\,\text{GeV}$ and $|\eta|<2.5$. 
Among the remaining photons we require at least two identified 
ones, using the procedure described in Sec.\ \ref{sec:results:aaa}, 
only changing the transverse momentum requirements to 
$\pT>20\,\text{GeV}$ for both the leading and the subleading 
identified photon. 
We further demand the angular separation of both photons 
to be $\Delta R(\gamma_1,\gamma_2)>0.4$ and each photon and 
the lepton to be $\Delta R(\ell,\gamma)>0.7$.
Furthermore we require the transverse mass of the lepton-neutrino system
to be larger than $40\, \text{GeV}$.
The inclusive cross sections and correction factors were 
detailed in Tab.\ \ref{tab:xsec}.

\begin{figure}[t!]
  \centering
  \includegraphics[width=0.32\textwidth]{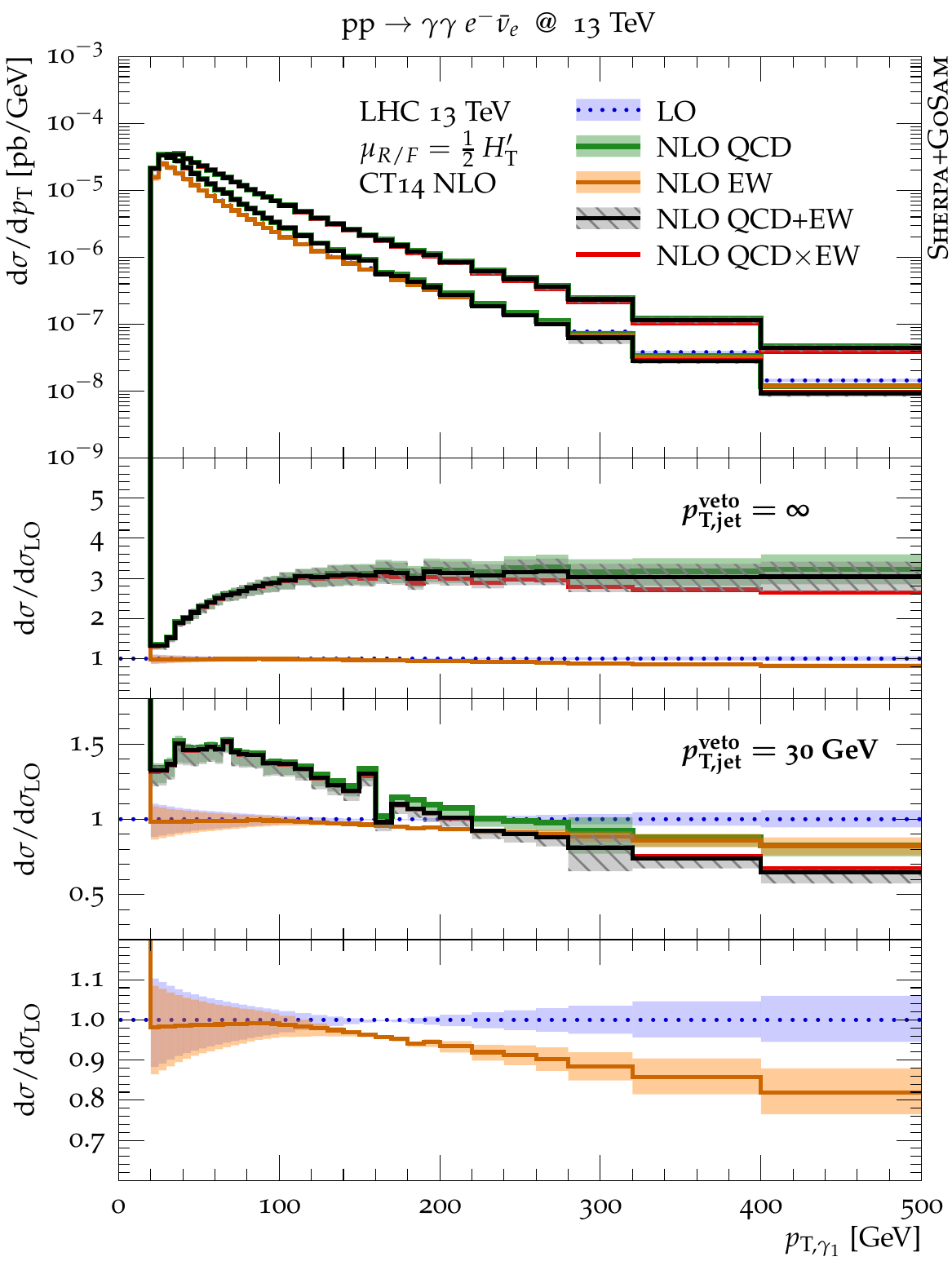}
  \includegraphics[width=0.32\textwidth]{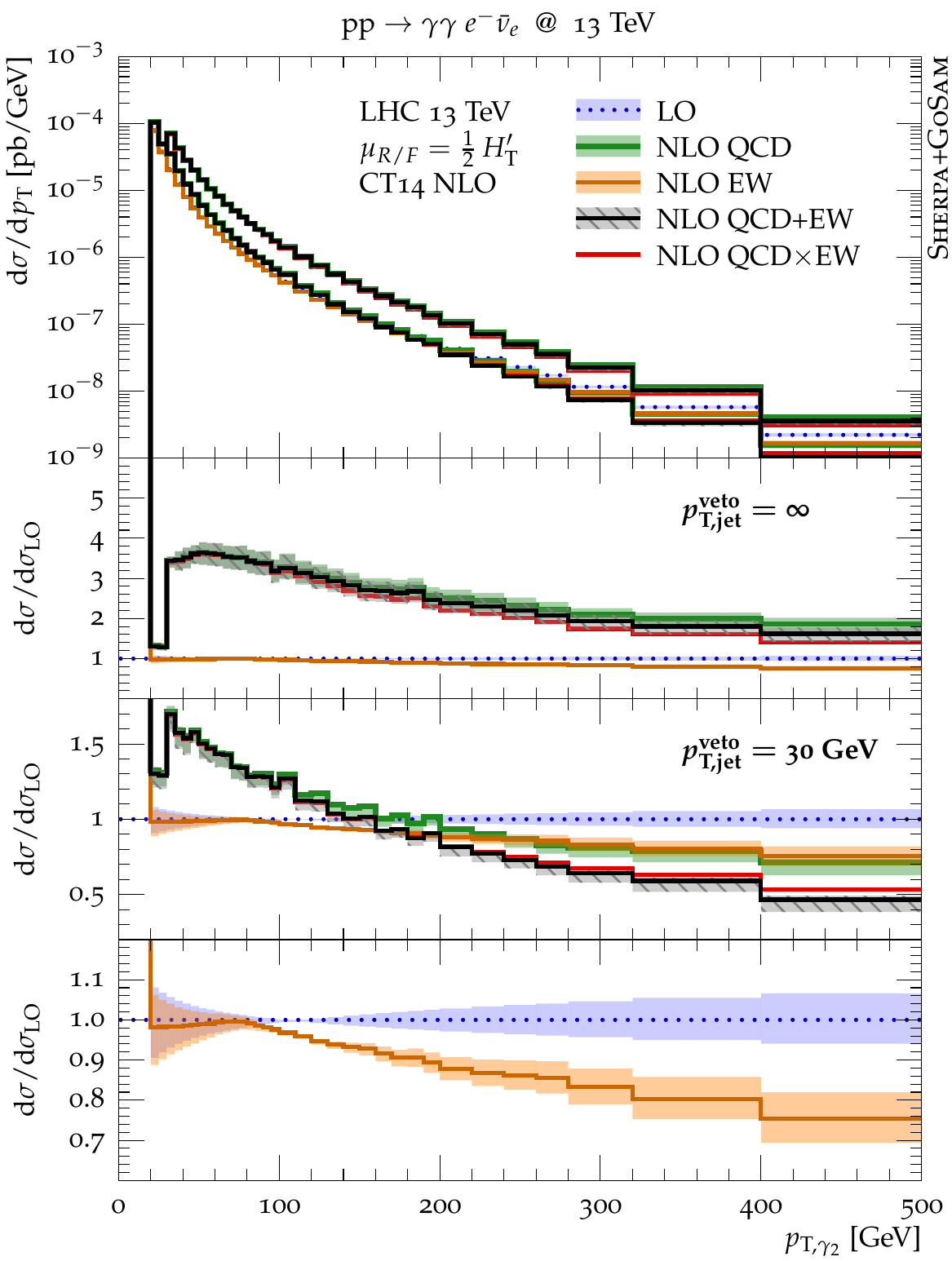}
  \includegraphics[width=0.32\textwidth]{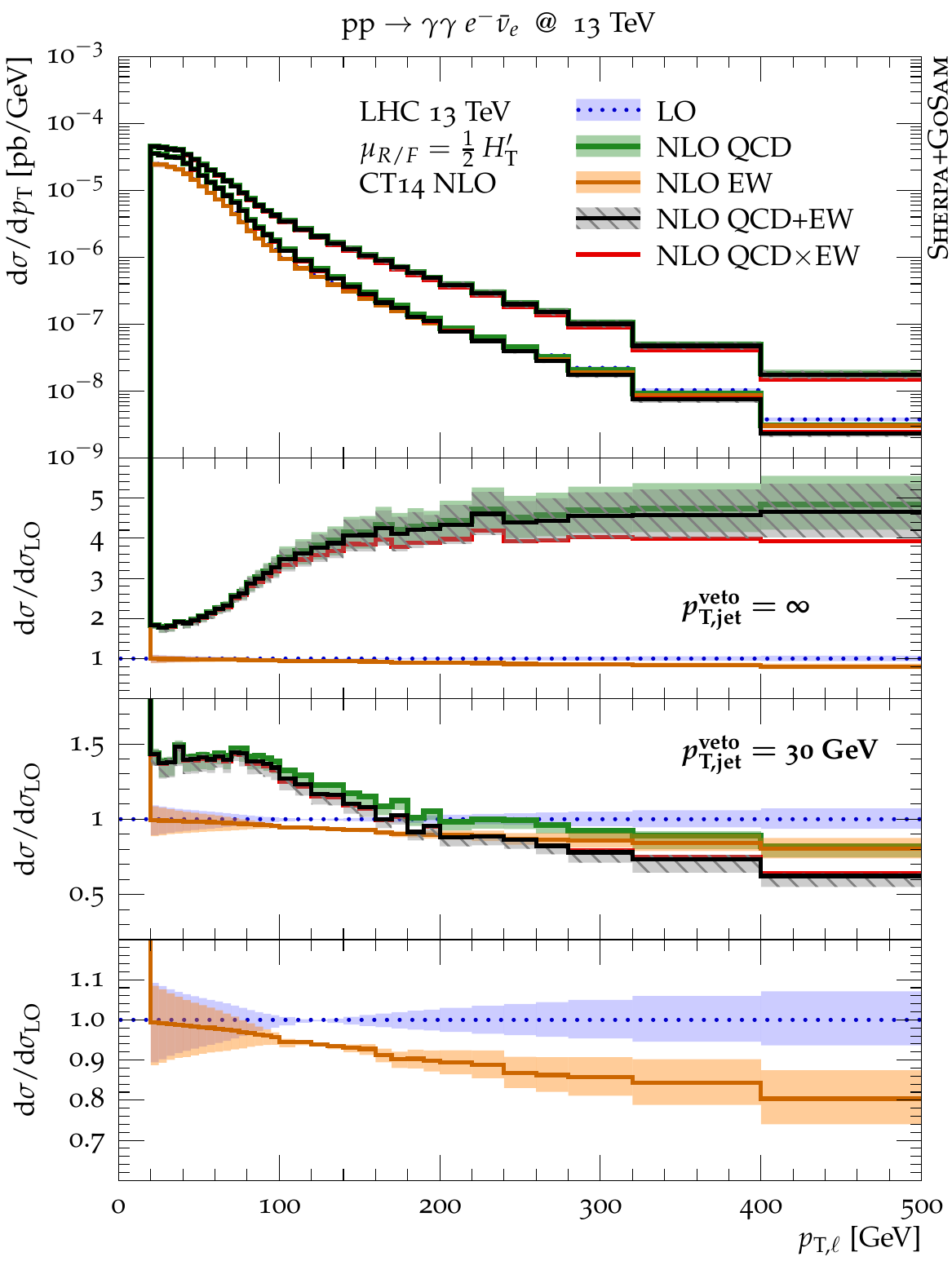}
  \caption{
    Transverse momentum of the leading (left), subleading (centre) 
    and third leading (right) photon at the LHC at 13\,TeV.
    \label{fig:aaw:pt}
  }
\end{figure}

Figure \ref{fig:aaw:pt} shows the transverse momenta of 
the leading and subleading photon as well as the charged 
lepton. 
As was observed in triple photon production, the inclusive 
NLO QCD corrections are large. 
Nonetheless, they differ substantially between the three 
observables and in their different regions. 
While $\deltaQCD$ rises from values of 0.3 at small 
transverse momenta of the leading photon to 2 at 
$\pT\approx 100\,\text{GeV}$, it remains constant above that 
value.
In the subleading photon's transverse momentum, the 
situation is different. 
Experiencing an acceptance cut induced jump from 0.3 to 2.5 at low \pT, 
it decreases steadily thereafter to reach $\deltaQCD=0.8$ 
at 500\,GeV. 
The behaviour of the QCD corrections to the lepton \pT\ are 
then again qualitatively similar to that of the leading 
photon, starting at $\deltaQCD=0.8$ at low \pT \, and rising 
steeply to 2.5 at 100\,GeV and then gradually leveling out 
at 3.5. 
Again, the presence of a restrictive jet veto has little 
effect on the small transverse momentum regions, but 
effectively contains the size of the NLO QCD corrections 
to remain below $\deltaQCD=0.5$. 
Instead, they are again driven negative, reaching 
$-20-30\%$ for all three observables. 

The electroweak corrections are negative throughout, 
reaching $-20\%$ for the leading photon and the lepton 
and $-25\%$ for the subleading photon at 500\,GeV. 
For this process, starting at transverse momenta of 
around 100\,GeV, the electroweak corrections are much 
larger than the LO uncertainty estimate.
Due to their larger size, as compared with the \aaa\ 
production process, the difference between their 
additive and multiplicative combination with the 
QCD corrections becomes more pronounced when no 
jet veto is applied. 
One reason is that the QCD corrections are dominated 
by real emission contributions, which only receive the 
not-so-small electroweak correction factor in the 
multiplicative combination, but not the additive one. 
While this correction factor is, strictly speaking, 
associated only with the Born configuration, its 
application to real-emission configurations is well 
motivated in the electroweak Sudakov-regime at 
large transverse momenta. 
The concurrence of the additive and multiplicative 
schemes in the presence of the jet veto now precisely 
originates in the taming of the QCD corrections by 
restricting the influence of real-emission topologies. 

\begin{figure}[t!]
  \centering
  \includegraphics[width=0.32\textwidth]{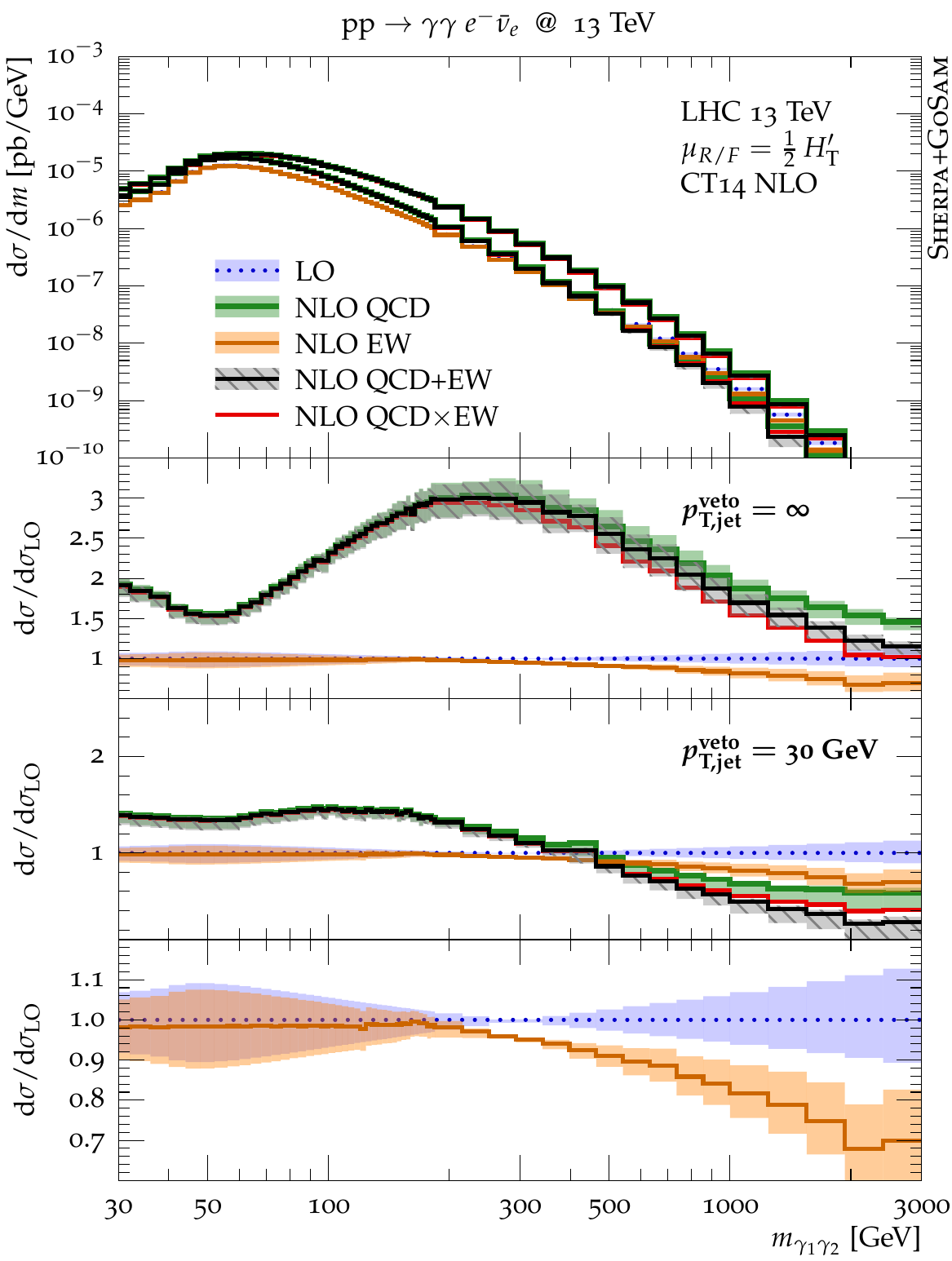}
  \includegraphics[width=0.32\textwidth]{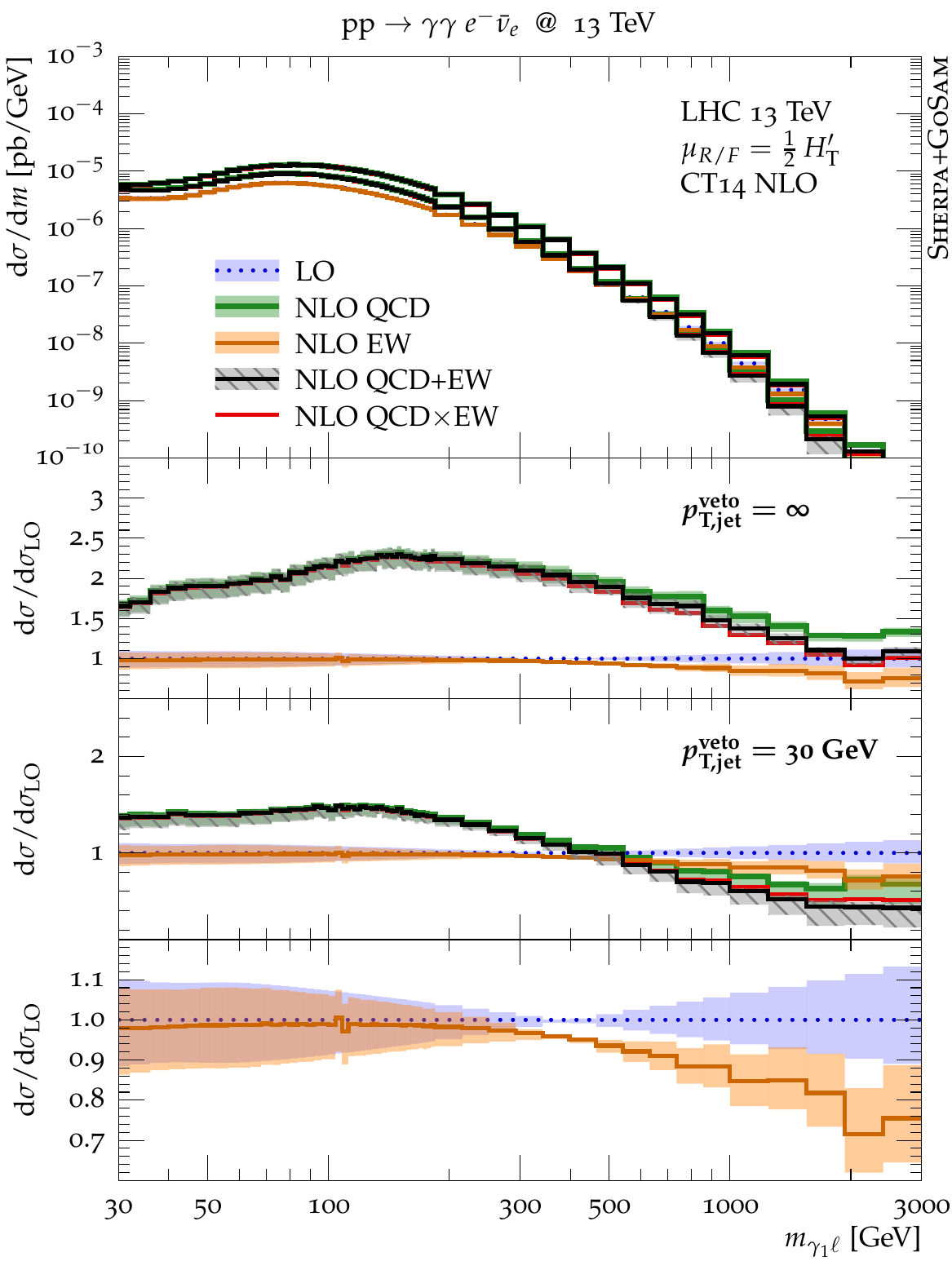}
  \includegraphics[width=0.32\textwidth]{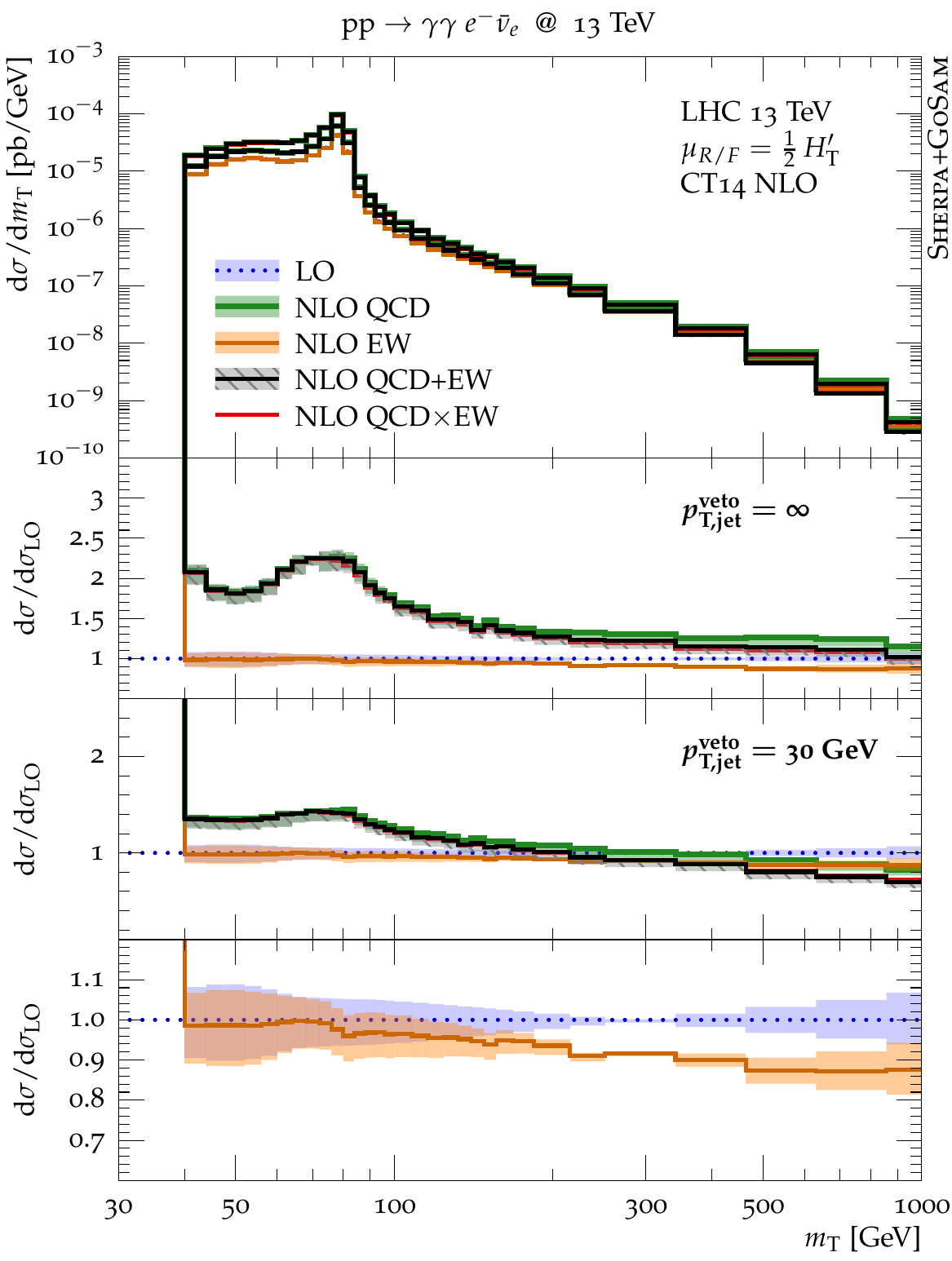}
  \caption{
    Pairwise invariant mass of the leading and subleading photon (left),
    leading and third leading photon (centre), subleading and third leading 
    photon (right) at the LHC at 13\,TeV.
    \label{fig:aaw:myy}
  }
\end{figure}

Considering now the invariant mass of both photons 
and the leading photon and the lepton, presented in 
Figure \ref{fig:aaw:myy}, a similar picture 
presents itself. 
Both QCD corrections are moderate at small invariant 
masses and increase as the invariant mass rises. 
While $m_{\gamma_1\gamma_2}$ develops a correction 
factor of $\deltaQCD=2$ at 200\,GeV and then falls 
again to 0.4 at 1\,TeV, $m_{\gamma_1\ell}$ only 
reaches a maximum of $\deltaQCD=1.3$ before falling 
to 0.4 as well. 
In the presence of a jet veto, the shape of the corrections 
remains similar, but their magnitude is contained to 
be smaller than 0.5, turning negative around 500\,GeV 
for both observables. 
The electroweak corrections are again negative 
throughout, reaching $-15-20\%$ at invariant masses 
of 1\,TeV.
Again, above 200\,GeV they are consistently outside 
the LO uncertainty estimate and play an important 
role.

Figure \ref{fig:aaw:myy} also displays the transverse 
mass of the $W$ boson on the right hand side. 
This distribution exhibits the general features known 
from inclusive $W$ production, only the QCD corrections 
are scaled to the values present in this process.
Electroweak corrections are less pronounced than for 
the invariant masses, but are still non-negligible. 
They reach about $-8\%$ at $m_\text{T}=200\,\text{GeV}$ 
and $-14\%$ at 1\,TeV.

\begin{figure}[t!]
  \centering
  \includegraphics[width=0.32\textwidth]{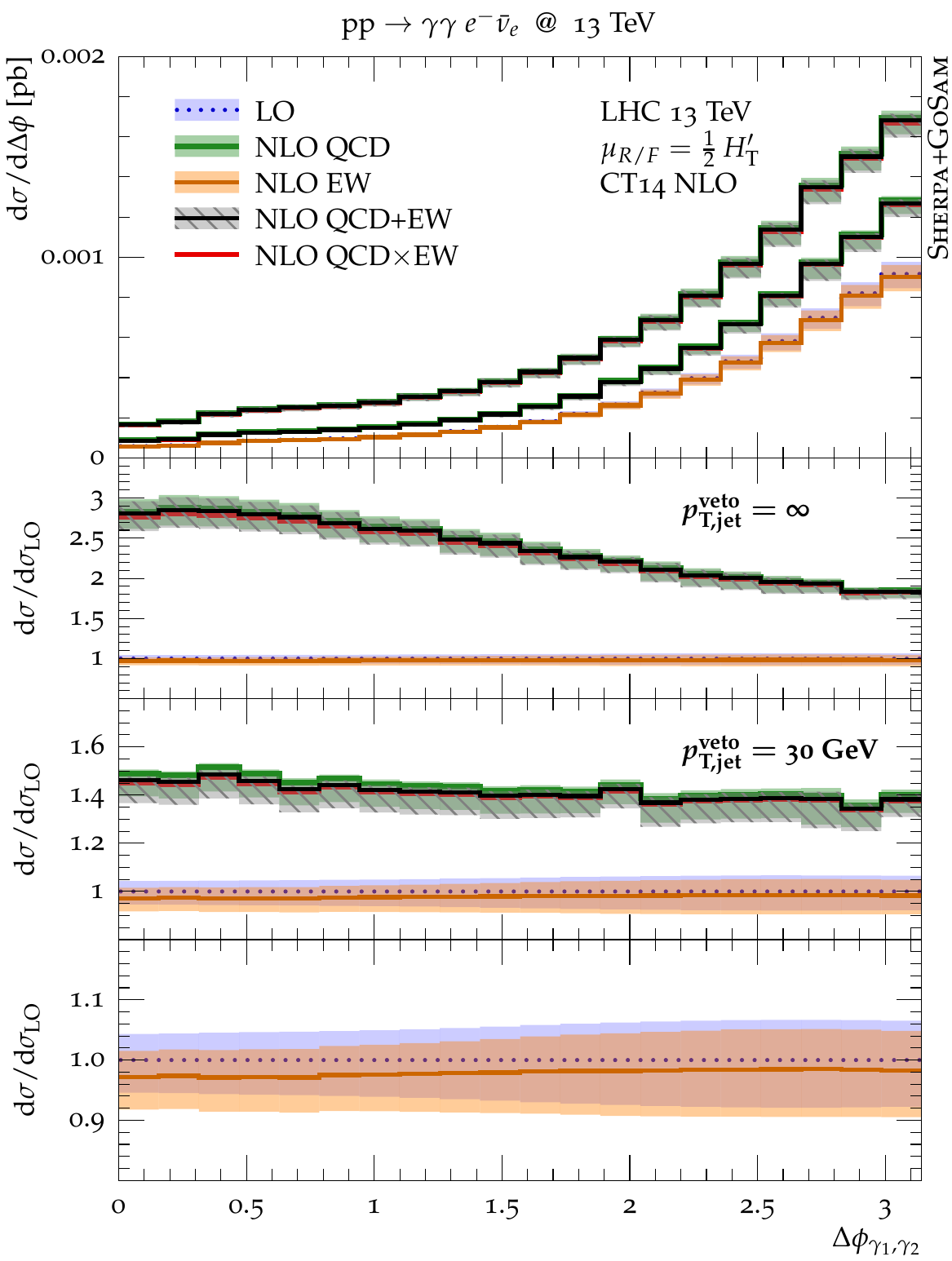}
  \includegraphics[width=0.32\textwidth]{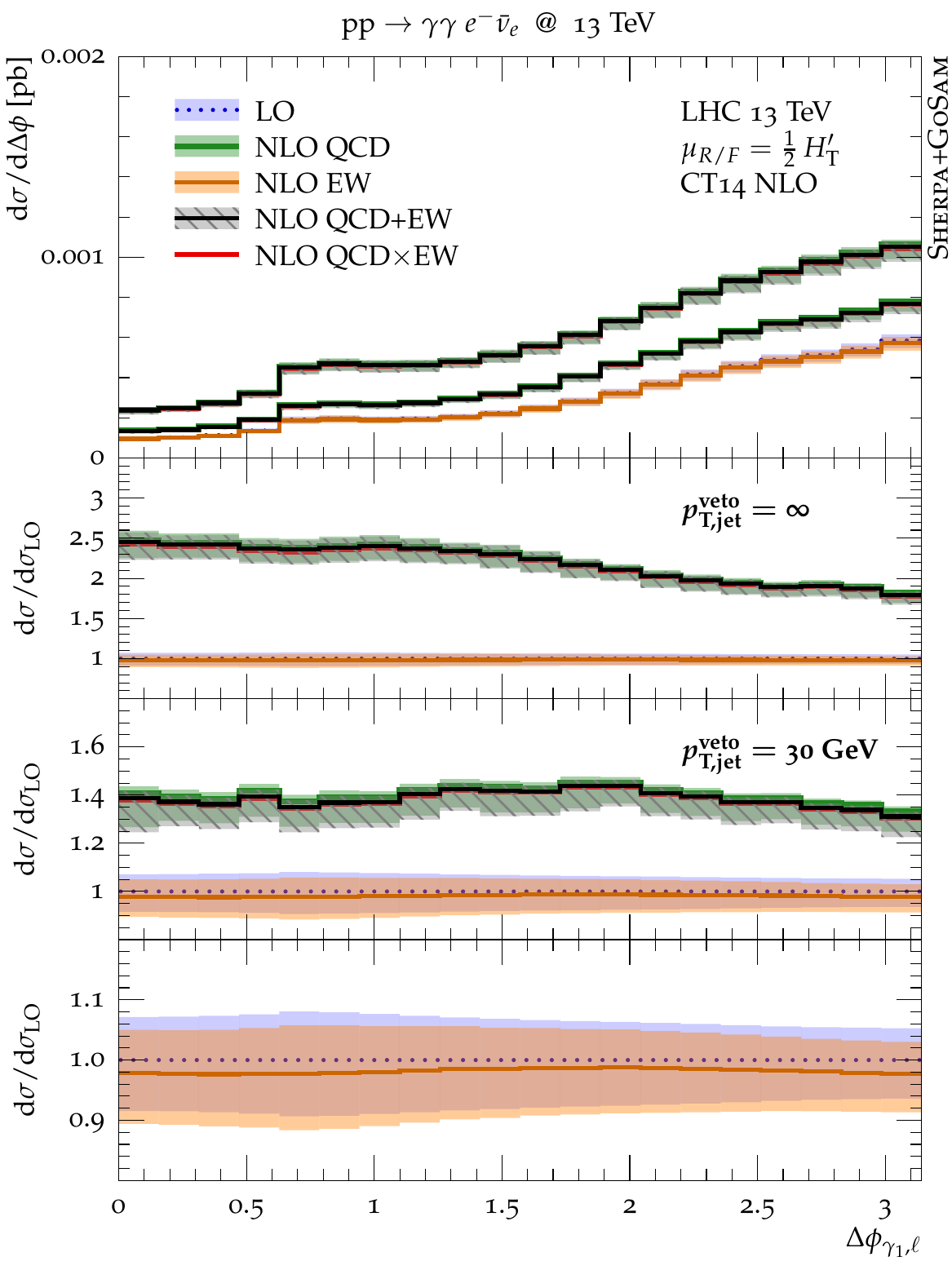}
  \includegraphics[width=0.32\textwidth]{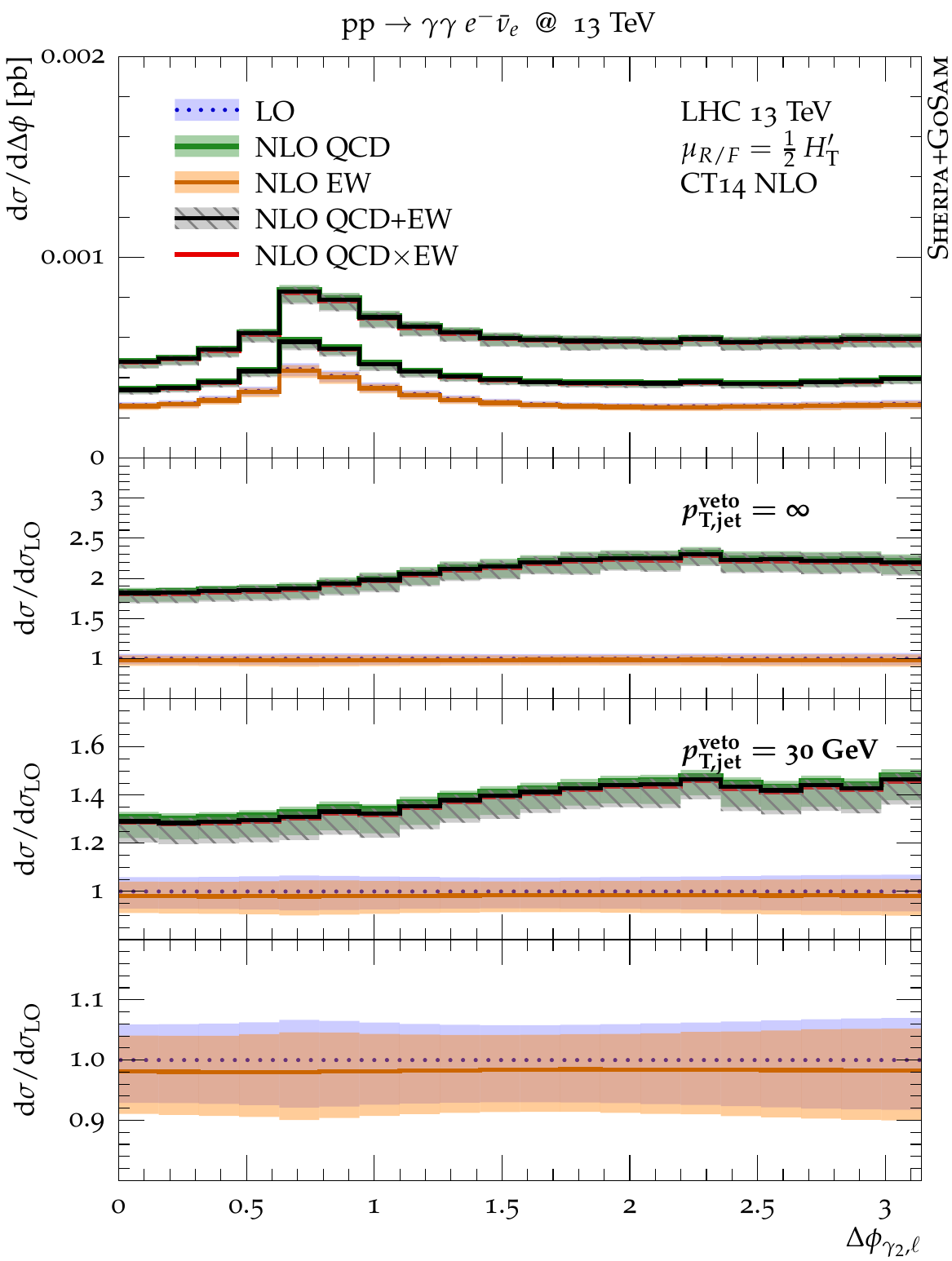}
  \caption{
    Azimuthal separation of the leading and subleading photon (left),
    leading and third leading photon (centre), subleading and third leading 
    photon (right) at the LHC at 13\,TeV.
    \label{fig:aaw:dphi}
  }
\end{figure}

Finally, Figure \ref{fig:aaw:dphi} presents the azimuthal 
separations of the leading and the subleading photon, and 
both the leading and subleading photon and the lepton. 
Contrary to \aaa\ production, no kinematic constraints 
are present at LO, owing to the difference in flavour 
between the considered objects and the fact that the 
lepton is only one of two decay products of the $W$ which 
takes the role of the third photon from point of view of 
the contributing diagrams. 
Both the QCD and EW corrections are generally flat, taking 
values of typically $\deltaQCD=1-2$ in absence and 
$\deltaQCD=0.4$ in presence of a jet veto, and 
$\deltaEW=-2\%$.

As none of the observables considered for this process 
exhibits a kinematic boundary at LO that is lifted at 
NLO, both the additive and the multiplicative combination 
of QCD and EW corrections present a viable prediction 
throughout and their difference can be taken as an 
indication of the potential size of higher-order corrections 
of $\order(\alphas\alpha)$.

\subsection[\texorpdfstring{$\gamma\gamma\ell^+\ell^-$}{aall} production]
           {$\boldsymbol{\gamma\gamma\ell^+\ell^-}$ production}
\label{sec:results:aaz}

Finally, we consider diphoton production in association with 
a $Z$ boson in its leptonic decay channel. 
In practise we look at lepton pair production including 
all non-resonant diagrams.
The fiducial phase space of our analysis is defined as follows, again according 
to existing measurements \cite{Aad:2016sau}. 
First, we require the presence of exactly one lepton pair 
of opposite charge, dressed with all photons in a cone of 
size $R=0.1$, with $\pT>20\,\text{GeV}$ and $|\eta|<2.47$. 
Their invariant mass must be larger than 40\,GeV. 
As in the case of \aaw production, we then require at least 
two identified photons among the remaining ones after dressing, 
using the procedure described in Sec.\ \ref{sec:results:aaa}, 
loosening the transverse momentum requirements to 
$\pT>15\,\text{GeV}$ for both the leading and the subleading 
identified photon. 
We further demand the angular separation of both photons 
to be $\Delta R(\gamma_1,\gamma_2)>0.4$ and each photon and 
the lepton to be $\Delta R(\ell,\gamma)>0.4$.
The inclusive cross sections and correction factors were 
detailed in Tab.\ \ref{tab:xsec}.

\begin{figure}[t!]
  \centering
  \includegraphics[width=0.32\textwidth]{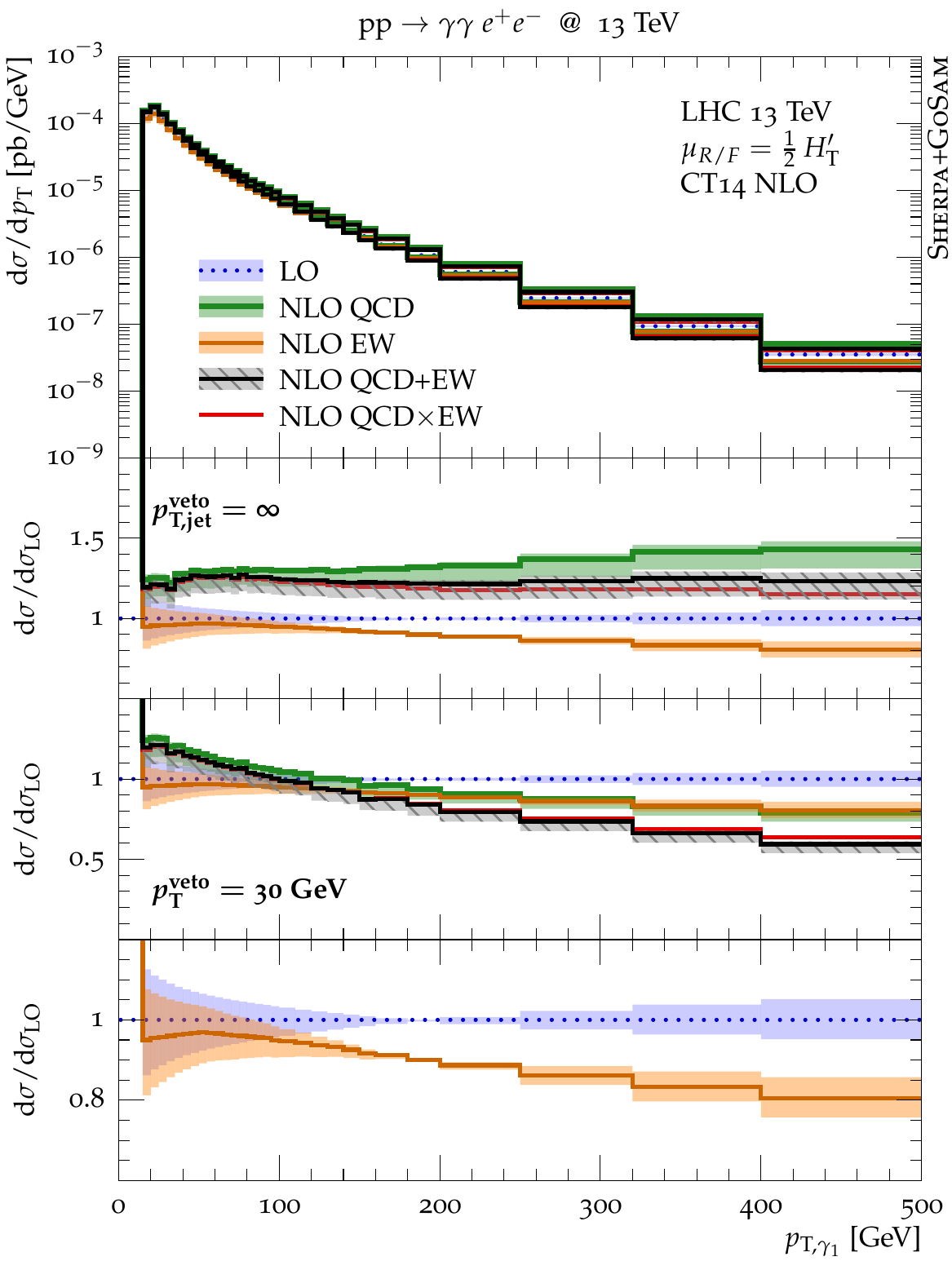}
  \includegraphics[width=0.32\textwidth]{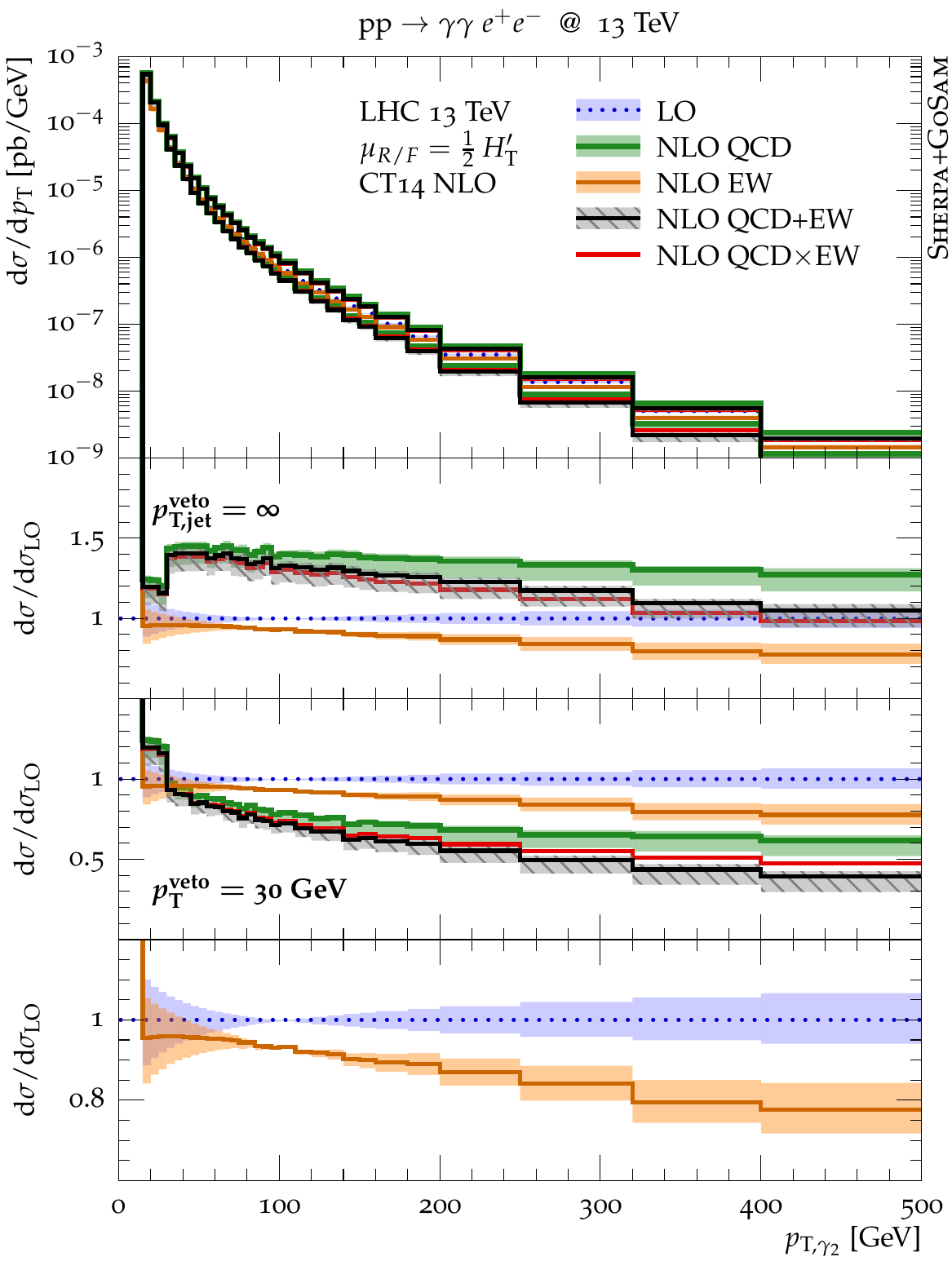}
  \includegraphics[width=0.32\textwidth]{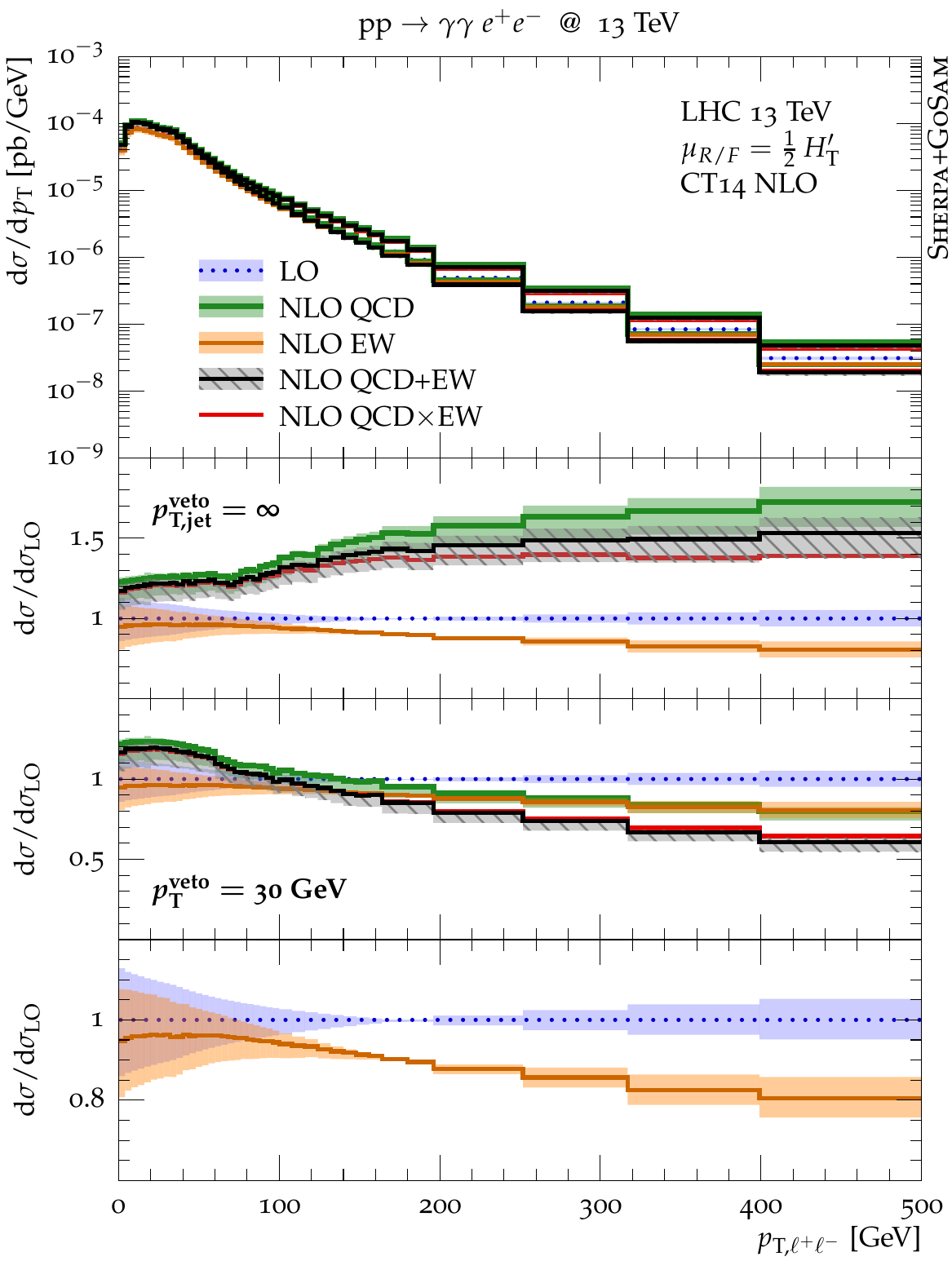}
  \caption{
    Transverse momentum of the leading (left) and subleading (centre) 
    photon as well as the dressed lepton pair (right) 
    in diphoton production in association with a lepton pair 
    at the LHC at 13\,TeV. 
    Details as in Fig.\ \ref{fig:aaa:pt}.
    \label{fig:aaz:pt}
  }
\end{figure}

In close analogy with the previous two processes considered, 
we start our discussion of the results with the transverse 
momenta of the leading and subleading photon and the charged 
lepton pair, cf.\ Figure \ref{fig:aaz:pt}. 
The first observation is, while the general behaviour of the 
spectra is the same as in \aaw production, the inclusive QCD 
corrections are much smaller in this case. 
They range between $\deltaQCD=0.3-0.4$ for the leading and 
subleading photon and rise to 0.7 for the lepton pair at 
500\,GeV. 
The restrictive jet veto employed for the previous process 
reduces the QCD corrections further. 
Beyond the negligible impact at very small transverse 
momenta, they turn negative early on, reaching $\deltaQCD=-0.2-0.4$ 
at 500\,GeV. 
The electroweak corrections are negative throughout, increasing 
from $\deltaEW=-4\%$ at low transverse momenta to $-20\%$ at 
500\,GeV.
They are consistently larger than the LO uncertainty estimate. 
As a consequence, the difference between the additive and 
multiplicative is comparably large, especially in the high 
\pT\ regions. 

\begin{figure}[t!]
  \centering
  \includegraphics[width=0.32\textwidth]{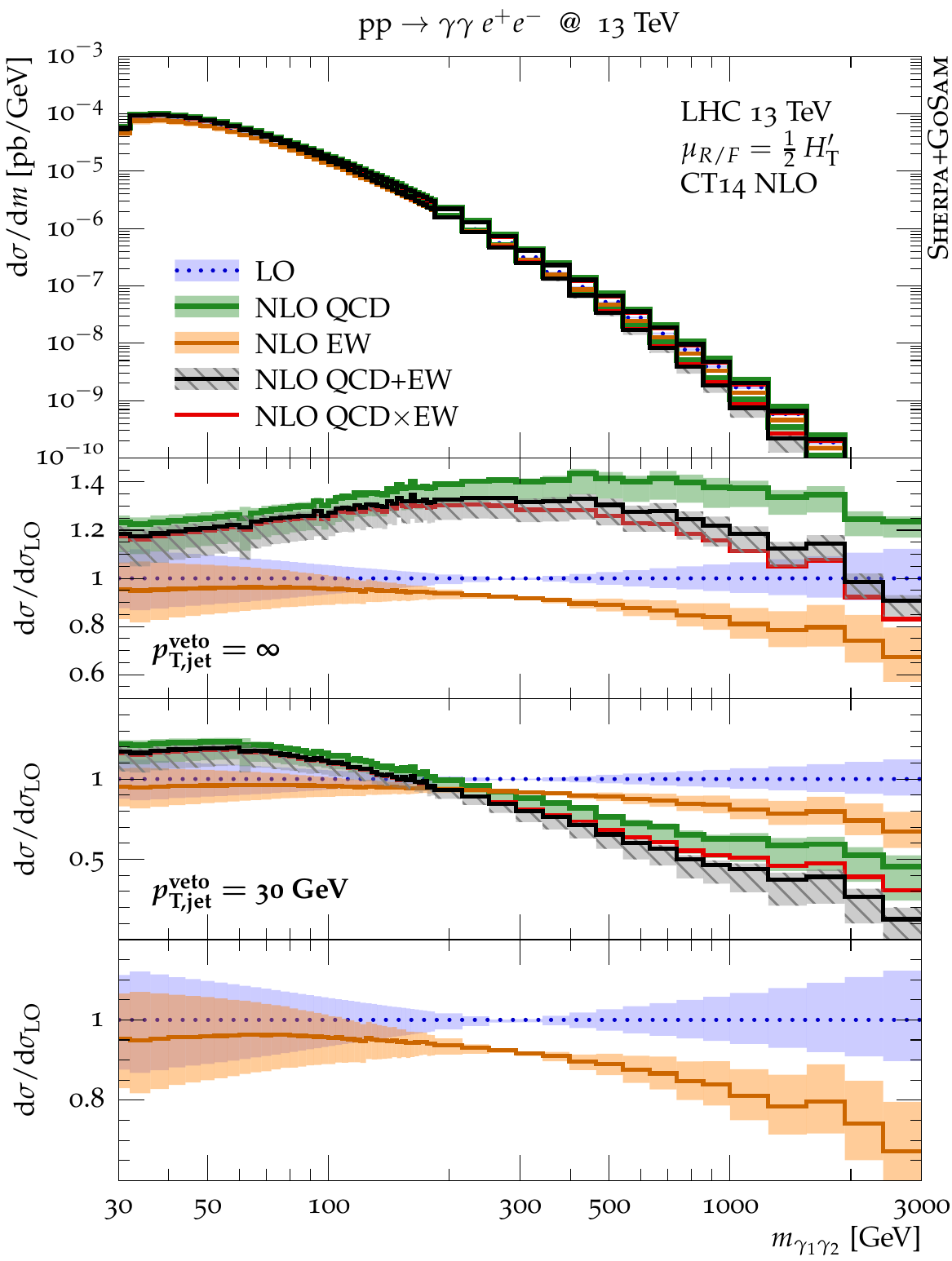}
  \includegraphics[width=0.32\textwidth]{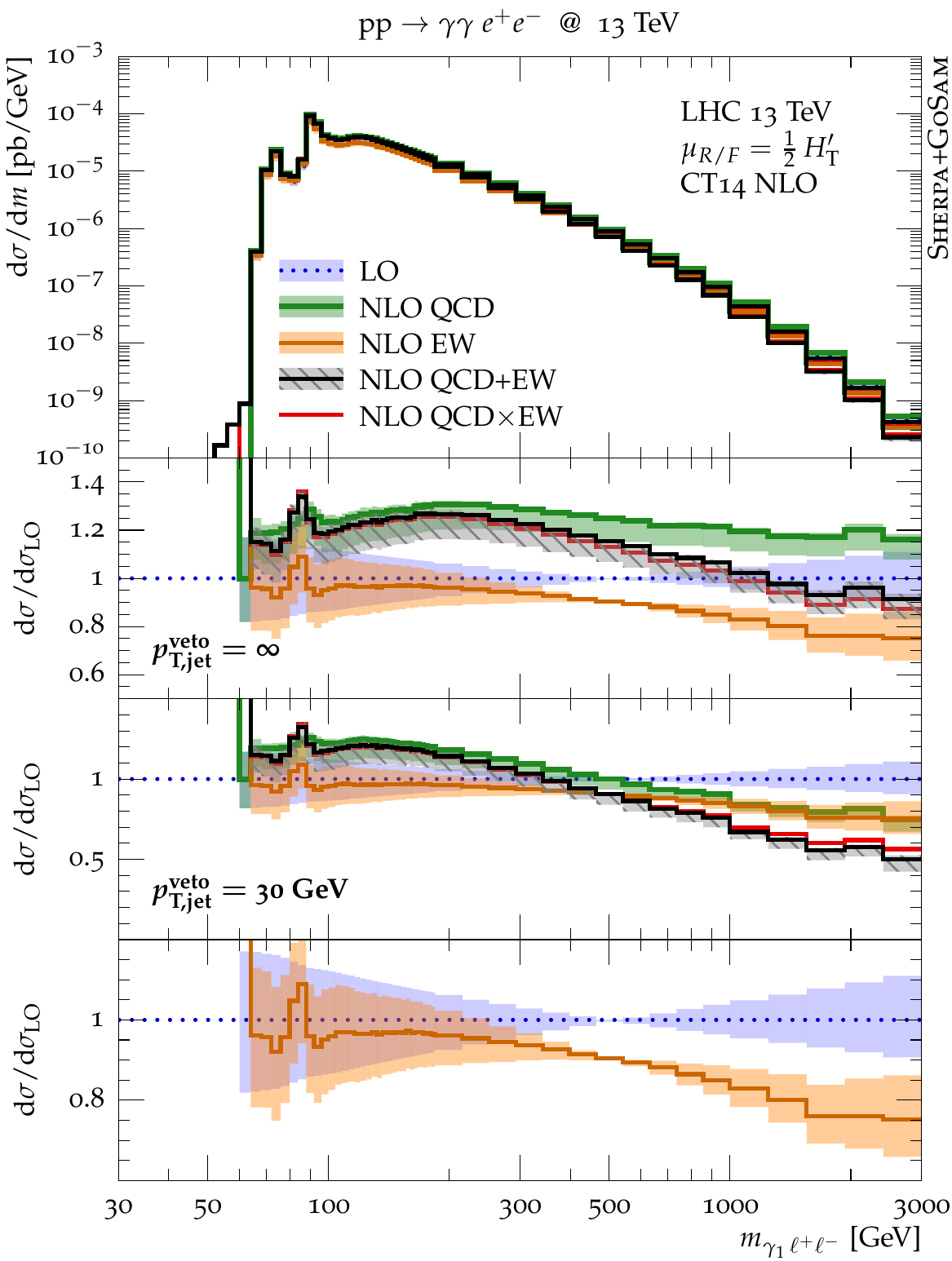}
  \includegraphics[width=0.32\textwidth]{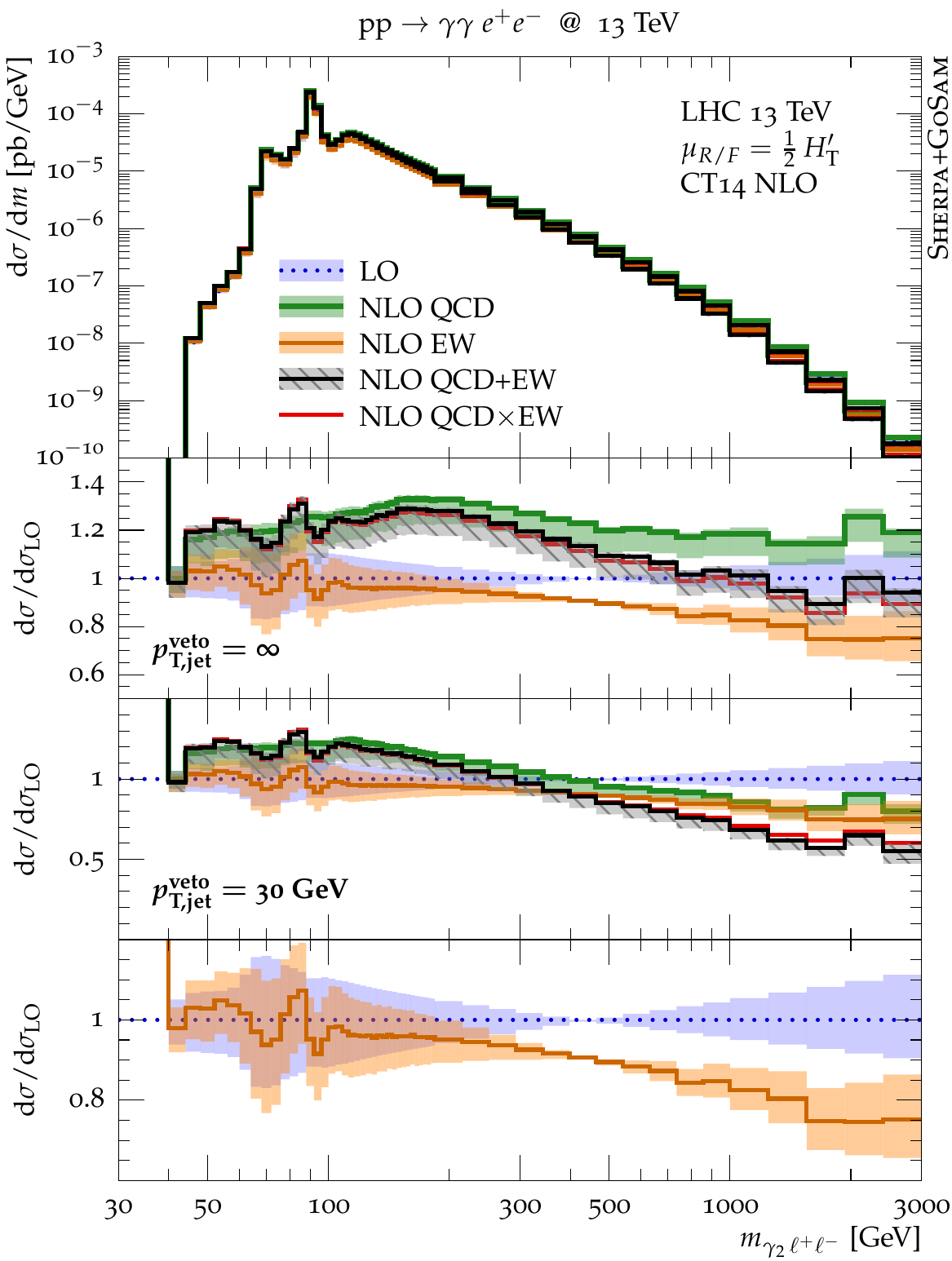}
  \caption{
    Pairwise invariant mass of the leading and subleading photon (left),
    the leading photon and the lepton pair (centre), and the subleading 
    photon and the lepton pair (right)
    in diphoton production in association with a lepton pair 
    at the LHC at 13\,TeV. 
    Details as in Fig.\ \ref{fig:aaa:pt}.
    \label{fig:aaz:myy}
  }
\end{figure}

Figure \ref{fig:aaz:myy} continues to display the invariant masses 
of the leading and subleading photon, and both the leading and 
subleading photon and the lepton pair. 
The familiar peak in $m_{\gamma_i\ell^+\ell^-}$ at the $Z$ boson mass 
is produced by 
configurations where the respective photon is emitted by the 
lepton-pair from a resonant $Z$ boson decay. 
Again, the QCD corrections are moderate, always staying below 
$\deltaQCD=0.4$ and of a similar shape as encountered already in 
\aaw\ production. 
The jet veto again reduces the QCD corrections and drives them negative 
in large portions of the observable range, reaching as much as 
$\deltaQCD=-0.5$ for the diphoton invariant mass. 
The electroweak corrections, on the other hand, are qualitatively very similar 
to the \aaw case: with the exception of the well known positive 
radiative corrections below the $Z$ peak, they are negative 
throughout, reaching $\deltaEW=-20\%$ at invariant masses of 
1\,TeV.
As in the case of the transverse momenta, due to the similar size 
of the QCD and electroweak corrections at large transverse 
momenta, the difference between their additive and the multiplicative 
combination is amplified.

\begin{figure}[t!]
  \centering
  \includegraphics[width=0.32\textwidth]{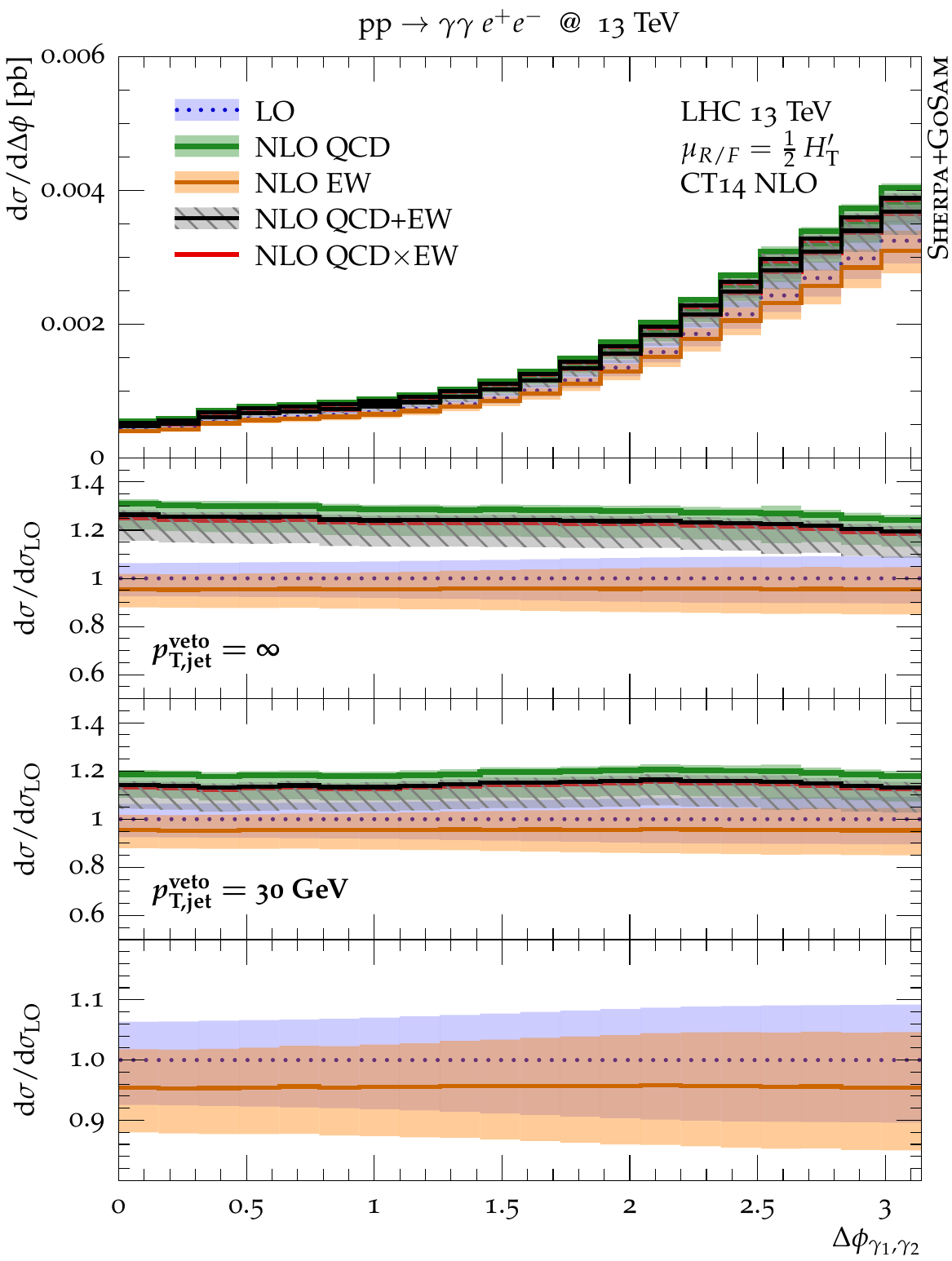}
  \includegraphics[width=0.32\textwidth]{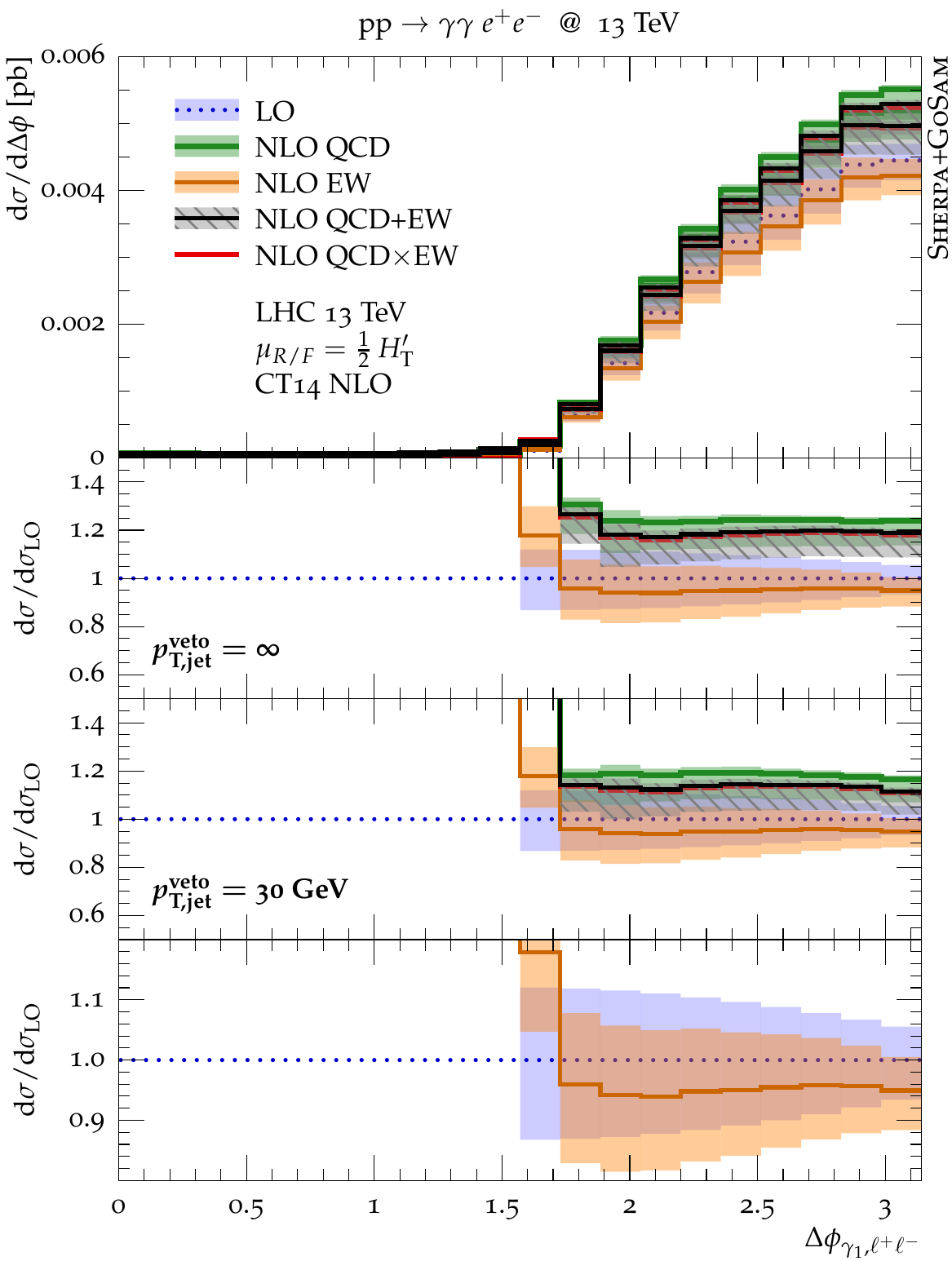}
  \includegraphics[width=0.32\textwidth]{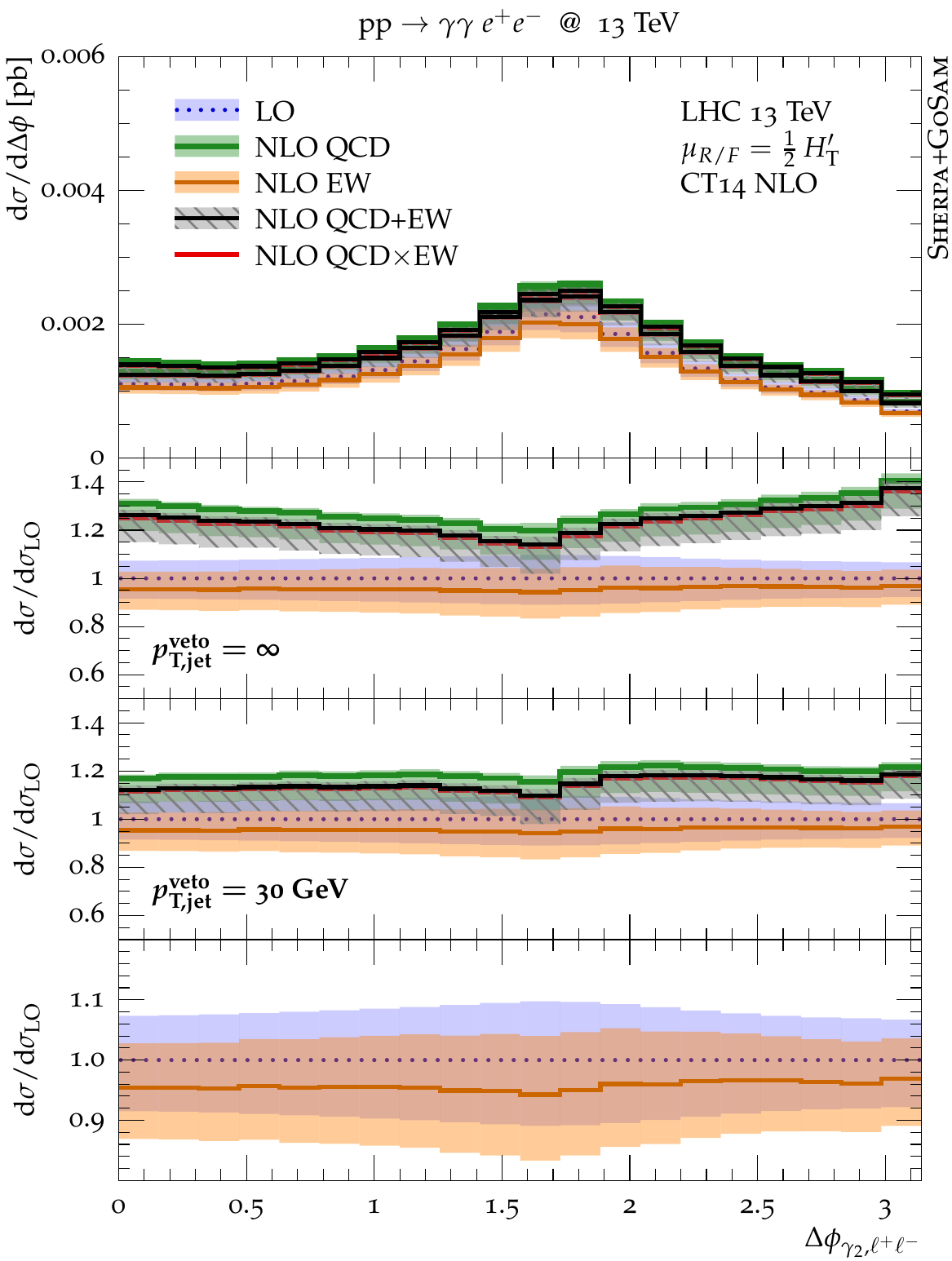}
  \caption{
    Azimuthal separation of the leading and subleading photon (left),
    the leading photon and the lepton pair (centre), and the subleading 
    photon and the lepton pair (right)
    in diphoton production in association with a lepton pair 
    at the LHC at 13\,TeV. 
    Details as in Fig.\ \ref{fig:aaa:pt}.
    \label{fig:aaz:dphi}
  }
\end{figure}

In Figure \ref{fig:aaz:dphi} we consider the azimuthal separation 
between the leading and subleading photon, and both the leading 
and subleading photon and the lepton pair.
While both photons tend to be back-to-back, the azimuthal 
separation of the leading photon and the lepton pair exhibits 
a kinematic edge at LO, restricting it to be larger than 
$\tfrac{1}{2}\,\pi$. 
As discussed above this kinematic edge is also present in the case of the triple photon process
(see Fig. \ref{fig:aaa:dphi}) when considering the azimuthal angle between
the leading and any subleading photon.
On the contrary, the azimuthal separation of the subleading 
photon and the lepton pair exhibits no such limit at LO and 
instead peaks at roughly $\Delta\phi=\tfrac{1}{2}\,\pi$. 
The QCD corrections, with the exception of the aforementioned 
edge, are generally flat, taking values of $\deltaQCD\approx 0.3$ 
in the absence and $\deltaQCD\approx 0.2$ in the presence 
of a jet veto. 
The electroweak corrections are equally flat and amount to 
$\deltaEW\approx-4\%$. 
Their additive and multiplicative combinations are very 
similar.

\begin{figure}[t!]
  \setlength{\unitlength}{\textwidth}
  \begin{picture}(0,0.37)
    \put(0,0.24){\includegraphics[width=0.32\textwidth]{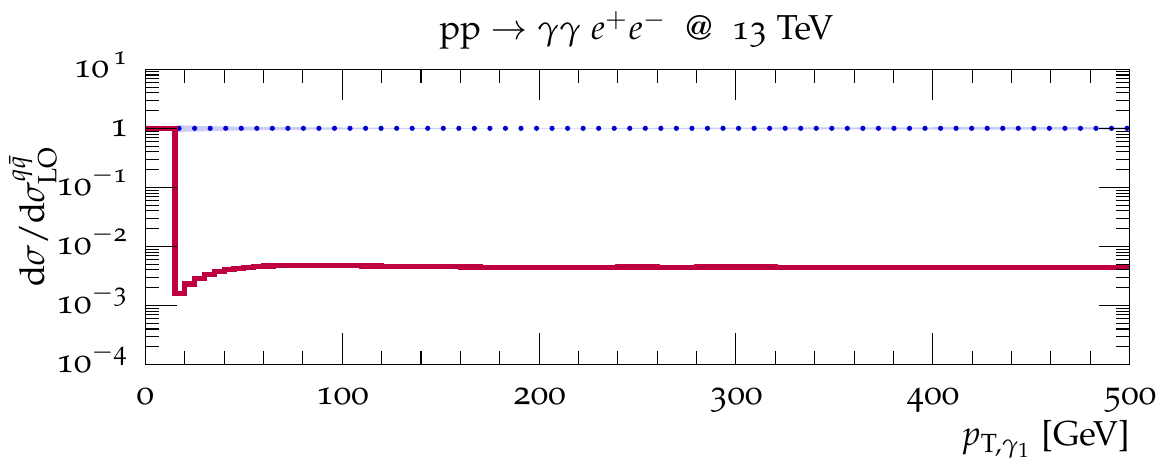}}
    \put(0,0.12){\includegraphics[width=0.32\textwidth]{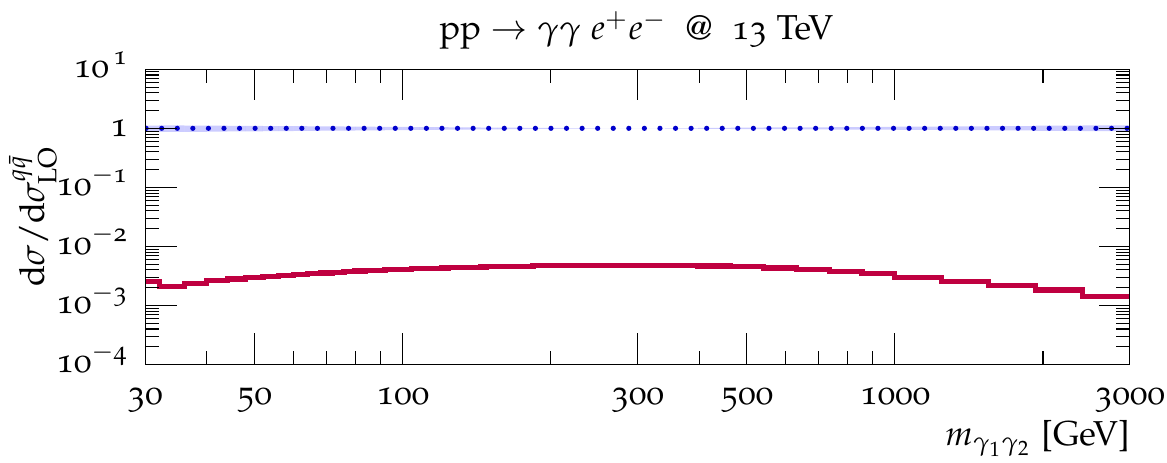}}
    \put(0,0){\includegraphics[width=0.32\textwidth]{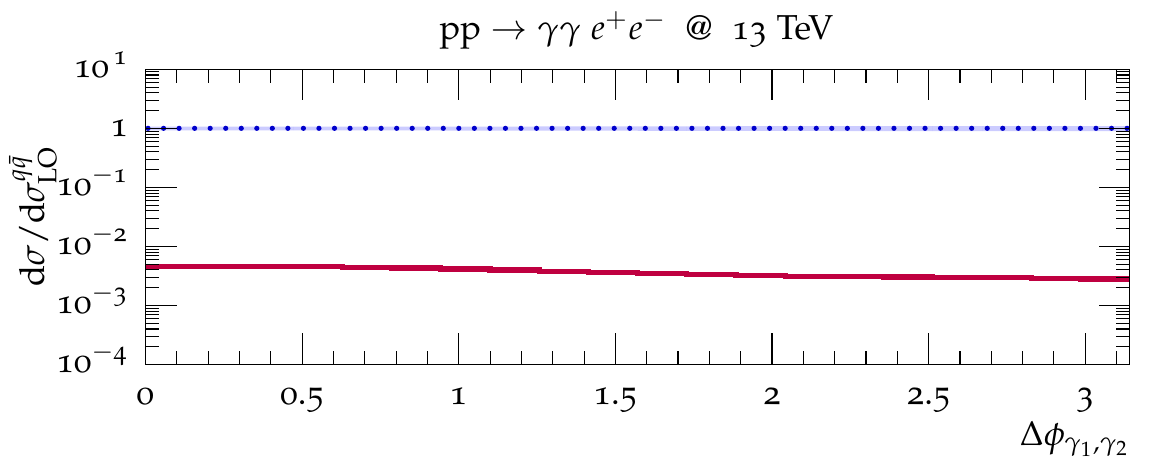}}
    \put(0.33,0.24){\includegraphics[width=0.32\textwidth]{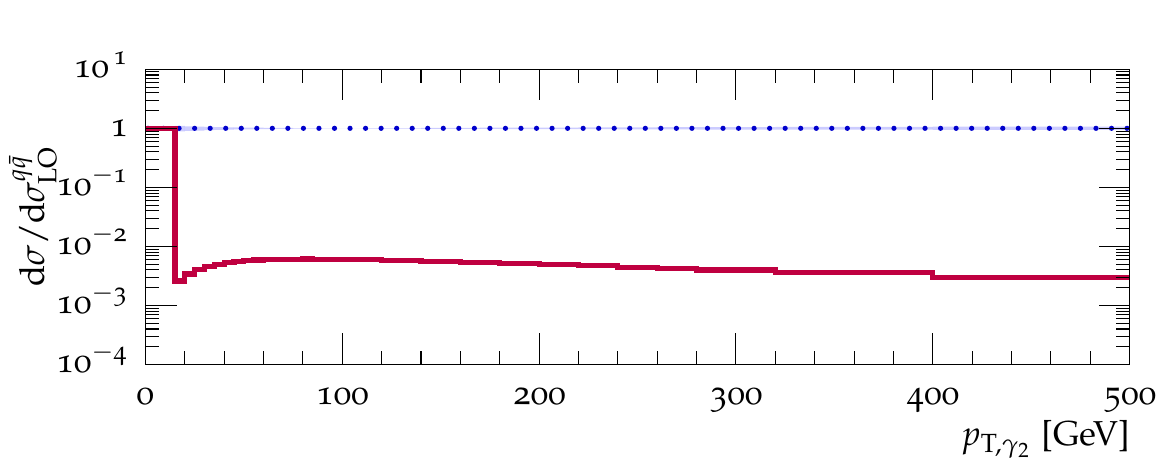}}
    \put(0.33,0.12){\includegraphics[width=0.32\textwidth]{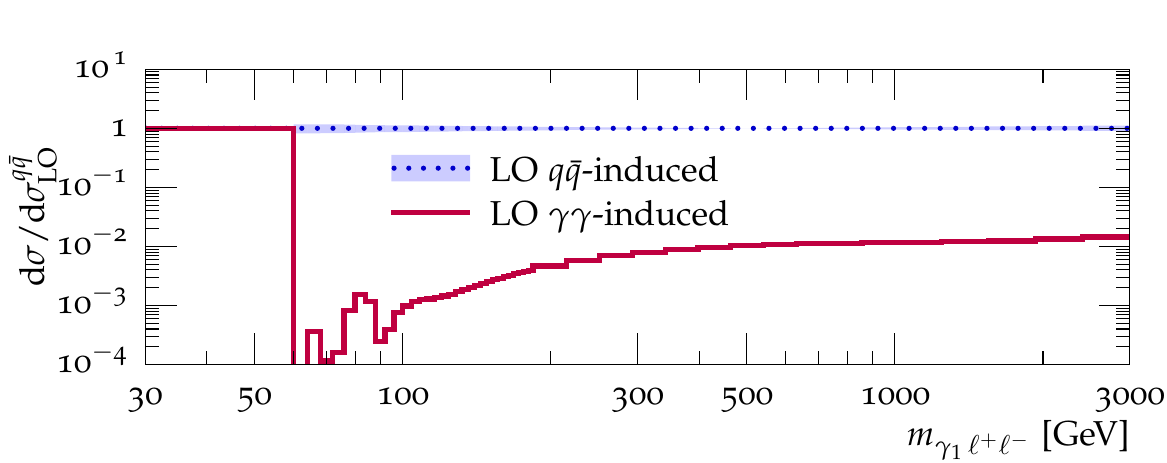}}
    \put(0.33,0){\includegraphics[width=0.32\textwidth]{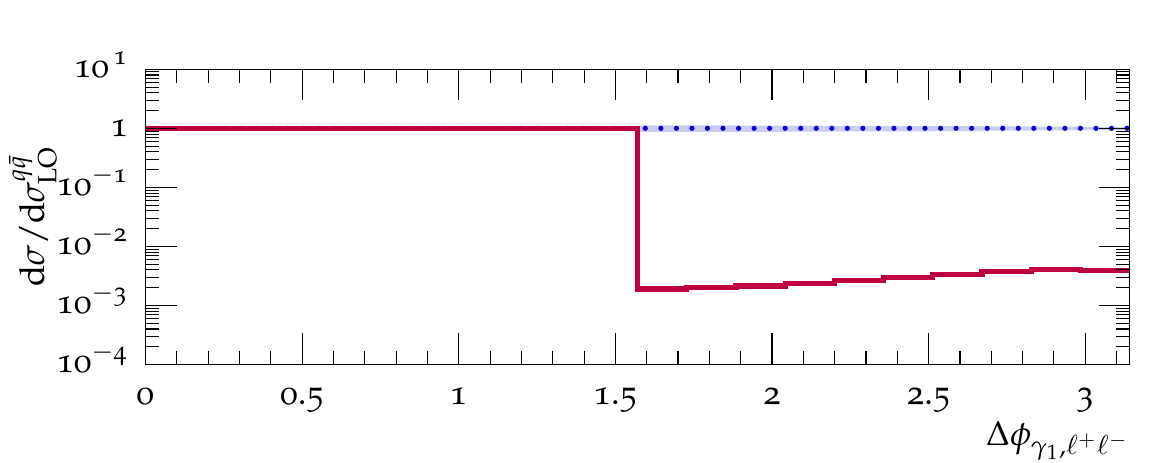}}
    \put(0.66,0.24){\includegraphics[width=0.32\textwidth]{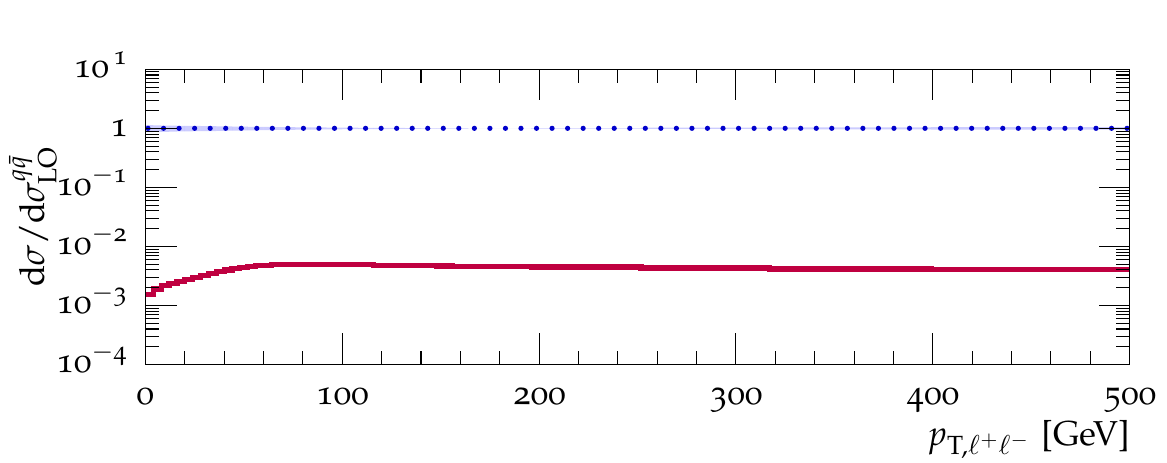}}
    \put(0.66,0.12){\includegraphics[width=0.32\textwidth]{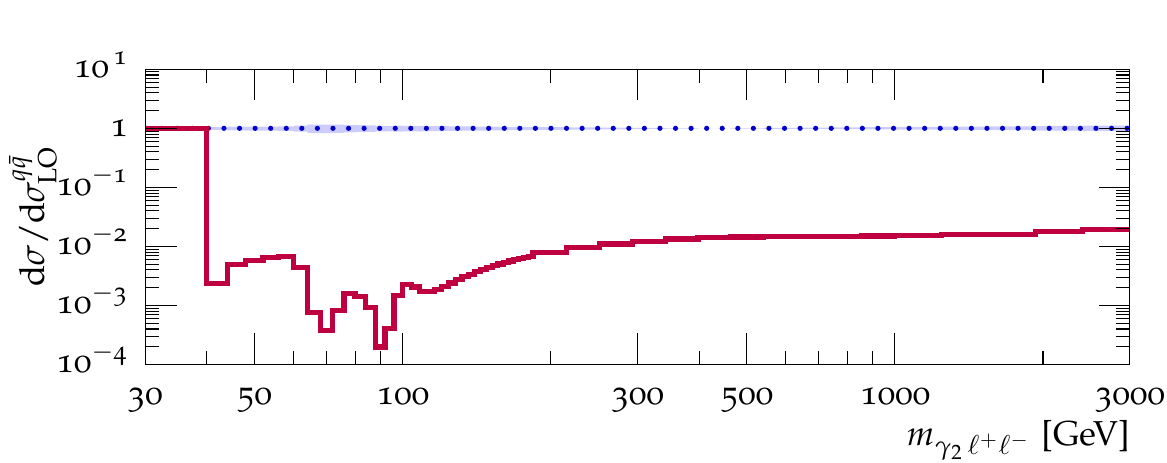}}
    \put(0.66,0){\includegraphics[width=0.32\textwidth]{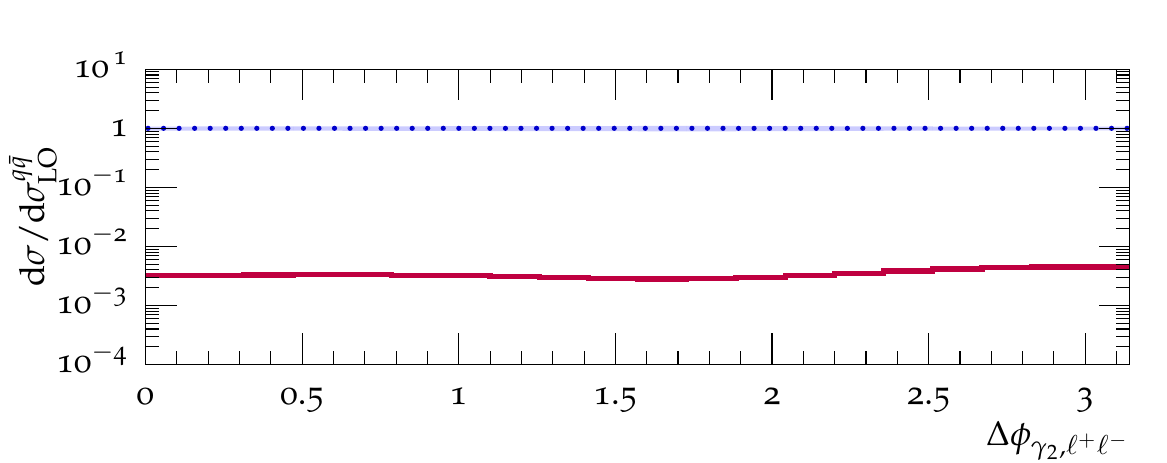}}
  \end{picture}
  \caption{
    Contribution of $\gamma\gamma$-induced production channels at LO.
    \label{fig:aaz:aa-ind}
  }
\end{figure}

Finally, \aaz\ production is the only process of the three processes 
considered in this publication which can be produced through 
photon-induced channels at leading order, 
$\gamma\gamma\to\ell^+\ell^-\gamma\gamma$. 
To this end, we investigate their contribution at LO in 
Figure \ref{fig:aaz:aa-ind}. 
The photon-induced contribution is typically a few per mille of the 
LO cross section, but raises up to 2\% for invariant masses larger 
than 1\,TeV, especially $m_{\gamma_i\ell^+\ell^-}$. 
In this region, however, typical QCD and EW corrections are much larger 
such that this contribution can be safely ignored. 
As the photon-induced processes only contribute through non-resonant 
lepton-pair production, their contribution can be further suppressed 
by tightening the acceptance window in the lepton-pair invariant mass.

\section{Conclusions}
\label{sec:conclusions}
Processes that involve vertices of three and four electroweak gauge bosons are among the most promising processes
where new physics might be found. Deviations from the Standard Model couplings or vertices that do not exist in the 
Standard Model might be found. The deviations from the Standard Model predictions can for instance conveniently be 
described by higher dimensional operators in terms of an effective field theory. Precise measurements of these  
processes allow to constrain and set limits to these higher dimensional operators.

In this paper we investigated the Standard Model predictions to a subset of such processes where we require
two photons plus an additional electroweak vector boson in the final state.  As additional vector boson we allowed
for a third photon as well as for a $W$ or $Z$ boson where we considered their leptonic decay modes.
This particularly implies that all off-shell and non-resonant contributions are taken into account.

We calculated the next-to-leading order QCD and electroweak corrections to these three processes for a set
of realistic fiducial cuts. Particular
emphasis has been put on the up to now unknown electroweak corrections as well as the combination of QCD and
electroweak corrections with the aim of producing the most precise prediction possible within the Standard Model.
As expected, we found the QCD corrections to be large and dominant compared to the electroweak corrections when 
calculating the corrections to the total cross sections. QCD corrections are particular large for these processes due to new
channels opening up at NLO but can effectively be reduced by applying a jet veto. Electroweak corrections
lead to moderate corrections to the total cross section of up to $4.4\%$ for the $\gamma\gamma \ell^{+}\ell^{-}$ 
process. However, in the high energy tail of differential distributions the electroweak corrections become increasingly important
and can lead to corrections of up to $30 \%$. Electroweak corrections become important in the same regions where one
also expects increasing effects of higher dimensional operators. The precise determination of limits on higher dimensional 
operators therefore requires the inclusion of higher order correction of both QCD as well as of the electroweak interaction.

We also compared the two possibilities of combining QCD with electroweak corrections, namely either in an additive or
in a multiplicative way. One expects that the difference between the two schemes is small as they can be seen as an estimation
of neglected higher order contributions. We found that although this is true on the level of total cross sections there can be 
regions in phase space where the leading order contribution vanishes and therefore one or both $K$-factors become infinitely large. 
For those kinematical edges we observe a breakdown of the multiplicative scheme whereas the additive scheme provides
a reliable estimation of the higher order uncertainties throughout the whole phase space.

\section*{Acknowledgements}
N.G.\ was supported by the Swiss National Science Foundation under contract
PZ00P2\_154829. M.S.\ was supported by PITN--GA--2012--315877 ({\it MCnet}) 
and the ERC Advanced Grant MC@NNLO (340983).


\bibliographystyle{amsunsrt_mod}
\bibliography{journal}

\end{document}